\documentclass[aos,preprint]{imsart}

\usepackage{amsthm,amsmath,natbib,color}
\usepackage{amssymb}
\usepackage{latexsym}
\usepackage{float}
\usepackage{graphicx}
\usepackage{amsfonts}
\usepackage{bbm}
\usepackage{longtable}
\usepackage{booktabs,url}

\arxiv{1702.07283}

\startlocaldefs

\def\ds{\displaystyle}

\def\R{\mathbb{R}}

\def\E{\text{E}}
\def\I{\text{I}}

\def\tr{\text{tr}}
\def\RSS{\text{RSS}}

\newtheorem{theorem}{Theorem}[section]
\newtheorem{lemma}[theorem]{Lemma}
\newtheorem{definition}[theorem]{Definition}
\newtheorem{condition}[theorem]{Condition}

\endlocaldefs

\begin{document}
\nocite{*}

\begin{frontmatter}

\title{Non-penalized variable selection in high-dimensional linear model settings via generalized fiducial inference}
\runtitle{GFI Variable Selection}


\author{\fnms{Jonathan P} \snm{Williams}\ead[label=e1]{jpwill@live.unc.edu}}
\address{\printead{e1}}
\and
\author{\fnms{Jan} \snm{Hannig}\ead[label=e2]{jan.hannig@unc.edu}}
\affiliation{University of North Carolina at Chapel Hill}

\runauthor{J. Williams and J. Hannig}

\begin{abstract}

Standard penalized methods of variable selection and parameter estimation rely on the magnitude of coefficient estimates to decide which variables to include in the final model.  However, coefficient estimates are unreliable when the design matrix is collinear.  To overcome this challenge an entirely new perspective on variable selection is presented within a generalized fiducial inference framework.  This new procedure is able to effectively account for linear dependencies among subsets of covariates in a high-dimensional setting where $p$ can grow almost exponentially in $n$, as well as in the classical setting where $p \le n$.  It is shown that the procedure very naturally assigns small probabilities to subsets of covariates which include redundancies by way of explicit $L_{0}$ minimization.  Furthermore, with a typical sparsity assumption, it is shown that the proposed method is consistent in the sense that the probability of the true sparse subset of covariates converges in probability to 1 as $n \to \infty$, or as $n \to \infty$ and $p \to \infty$.  Very reasonable conditions are needed, and little restriction is placed on the class of possible subsets of covariates to achieve this consistency result.

\end{abstract}



\end{frontmatter}

\section{Introduction}\label{Intro}

A strategy for developing variable selection procedures with desirable consistency properties entails exploiting some distinguishing property of the theoretical true data generating model.  For example, standard penalized methods of variable selection within a linear model framework such as LASSO of \cite{Tibshirani}, SCAD of \cite{FanandLi}, and the Dantzig Selector of \cite{Candes2007} rely on the magnitude of the coefficients in the true data generating model being relatively larger than those of the other coefficients.  \cite{Johnson2012} use this property to construct nonlocal prior densities over all subsets of covariates.  The defining property of their nonlocal density is that it takes the value of zero for subsets containing a covariate with a zero-valued coefficient.

We propose a more desirable way for eliminating redundancies from the sample space of candidate subsets which does not explicitly rely on coefficient magnitudes.  That is, any candidate true model should be non-redundant in the sense that it contains the minimal amount of information necessary for explaining and/or predicting the observed data.  One such criterion to exploit this non-redundancy property is that the only subsets with nonzero posterior probability should be those which cannot be predicted to some chosen precision by a subset of fewer covariates.  Such a criterion requires constructing a probability distribution on the space of candidate models, which is consistent with a Bayesian or fiducial variable selection paradigm.  The literature on high-dimensional linear models is vast, but we hope to contribute to it by using this setting to build a foundation for a fresh perspective on variable selection.

Recent work in the Bayesian high-dimensional linear model setting includes \cite{Rockova2015} who develop methods for separable and non-separable spike-and-slab penalized estimation, the credible set approach of \cite{Bondell2012}, and \cite{Narisetty2014} who propose a method based on shrinking and diffusing coefficient priors in which the variance of the priors are sample size dependent.  \cite{Lai2015} layout framework for penalized estimation within a GFI approach.  

\cite{Ghosh2015} provide insights into complications in Bayesian variable selection.  Namely, the size of the sample space ($2^{p}$) is often too large to compute all model probabilities, and even typically larger than can reasonably be sampled by Markov chain Monte Carol (MCMC) methods.  Thus, the nonlocal prior approach of \cite{Johnson2012} can achieve asymptotic consistency (where other approaches can only achieve pairwise consistency) because it is able to effectively eliminate a large enough portion of the $2^{p}$ subsets from the sample space.  To illustrate this point, consider the following simple example.  Let
\vspace{-.18cm}\begin{equation}\small\label{SimpleEx}
Y \sim \text{N}_{n}\Big(\beta_{1}\cdot x^{(1)} + \dots + \beta_{p}\cdot x^{(p)}, \sigma^{2} I_{n}\Big),
\vspace{-.18cm}\end{equation}
where $\beta_{j} \in \R$ and $x^{(j)} \in \R^{n}$ for $j \in \{1,\dots,p\}$, and $\sigma > 0$.  Further, suppose that the true but unknown values of $(\beta_{1}, \beta_{2}, \beta_{3}, \dots, \beta_{p})'$ are $(b_{1}, b_{2}, 0, \dots, 0)'$.  Within the nonlocal prior framework, the only subsets with non-negligible posterior probability are contained in the set $\big\{ \{x^{(1)} \}, \{x^{(2)} \}, \{x^{(1)} , x^{(2)} \} \big\}$.

When viewed as a prior density on the coefficients, nonlocal priors assign zero prior density to the true parameter value when the true parameter value is zero.  From a Bayesian perspective this is philosophically problematic, but very insightful for consistency of model selection.  The insight lends itself to the question: What other properties might the true model have which can be exploited to develop a statistical procedure with the ability to effectively eliminate subsets from the sample space?  

In addressing this question, we build our proposed methods from the idea that any candidate true model, as determined by the actual non-zero parameter values, should be non-redundant in the sense that it contains the minimal amount of information necessary for explaining and/or predicting the observed data.  We denote such subsets of the parameter space as $\varepsilon$-$admissible$, and define them precisely in Definition \ref{admissibility}.  Then, using the above nonlocal prior example, the entire model space $\{x^{(1)}, x^{(2)}, x^{(3)}\}$ for instance, is not $\varepsilon$-$admissible$  because it can be perfectly predicted by the smaller subset $\{x^{(1)}, x^{(2)}\}$.

To further illustrate the intuition behind our proposal, consider an example where $x^{(2)}$ is highly collinear with all of $x^{(3)}, \dots, x^{(p)}$ but is $not$ correlated with $x^{(1)}$, and that the true values of $(\beta_{1}, \beta_{2}, \beta_{3}, \dots, \beta_{p})'$ are $(b_{1}, b_{2}, b_{3}, \dots, b_{p})'$ with $b_{j} \ne 0$ for all $j \in \{1,\dots,p\}$.  In this case, assuming strong enough collinearity, $\exists c \in \R$ with $c\cdot x^{(2)} \approx b_{2}\cdot x^{(2)} + \dots + b_{p}\cdot x^{(p)}$, i.e.,
\vspace{-.18cm}\[\small
\big\| \big(b_{1}\cdot x^{(1)} + \cdots + b_{p}\cdot x^{(p)}\big) - \big(b_{1}\cdot x^{(1)} + c\cdot x^{(2)}\big) \big\| < \varepsilon
\vspace{-.18cm}\]
where $\varepsilon > 0$ is some desired precision.  Thus, for much of the parameter space the subset $\{x^{(1)}, \dots, x^{(p)}\}$ is not $\varepsilon$-$admissible$, but would be assigned nonzero posterior probability in the nonlocal prior framework.

We construct a posterior-like probability distribution over all subsets, which assigns negligible probability to elements that are not $\varepsilon$-$admissible$.  In constructing the posterior-like probability distribution we adopt a generalized fiducial inference (GFI) approach because it has similar to an objective Bayes interpretation with data driven priors, gives a systematic method of constructing a distribution function given a data generating equation such as a linear model, and it does not suffer from the issue of arbitrary normalizing constants which arise in many objective Bayesian priors \citep{Berger2001}.  In this manuscript we will provide a gentle introduction to GFI. A fuller account of GFI is given in the recent review paper \cite{Hannig2016}.

An advantage of both our approach and the nonlocal prior approach of \cite{Johnson2012} is that in addition to providing theoretical guarantees, our statistical procedures yield estimates of the posterior distribution over subsets of covariates.  This is in contrast to frequentist penalization based methods or Bayesian procedures fully dedicated to maximum a posteriori probability (MAP) estimation.  Such methods do not yield the posterior probability of a chosen model (i.e., the relative probability, given the observed data, of a given model against competing models).  Furthermore, \cite{Ghosh2015} argue that joint summaries of subsets of covariates are more robust to collinearity.  

With our approach to constructing a posterior-like distribution whose probability mass function value is negligible for subsets of the parameter space which are not $\varepsilon$-$admissible$, we are able to show that the probability of the true data generating model converges to 1 asymptotically in $n$ and $p$.  This consistency is shown to be true even with $p$ growing almost exponentially in $n$.  The reason being that the true model yields a stronger signal since it no longer has to compete within an overly redundant sample space.

The paper is organized as follows.  Section \ref{methodology} serves to introduce the general methodology and computational algorithm for carrying out our variable selection procedure based on a recent algorithm for explicit $L_{0}$ minimization \citep{Bertsimas2016}, which is fast enough to be used on real data.  The conditions needed for the main results, and the main results are presented and discussed in Section \ref{TheoryResults}.  Proofs are organized in the appendix.  We demonstrate the empirical performance of our procedure and compare it to other Bayesian and frequentist methods in simulation setups on synthetic data in Section \ref{SimResults}.  Computer code implementing our procedure is provided at \url{https://jonathanpw.github.io/software.html}.

\section{Methodology}\label{methodology}

As described in the previous section, our idea behind exploiting a non-redundancy property of the true data generating model relies on constructing a probability distribution concentrated on what we denote as $\varepsilon$-$admissible$ subsets.  This object is defined precisely in Definition \ref{admissibility}, but first an aside on the notation used throughout the paper.

Let $Y$ be an $n$-dimensional random vector, $X$ an $n\times p$ matrix with columns scaled to have unit norm, and $\beta^{0}$ a fixed $p$-dimensional vector with nonzero (or active) components indexed by the subset $M_{o} \subset \{1,\dots,p\}$, with 
\vspace{-.18cm}\begin{equation}\small\label{trueModel}
Y \sim \text{N}_{n}\Big(X_{M_{o}}\beta_{M_{o}}^{0}, (\sigma_{M_{o}}^{0})^{2}I_{n}\Big).
\vspace{-.18cm}\end{equation}
The design matrix denoted by $X_{M_{o}}$ is defined as the matrix composed of only those columns of $X$ corresponding to the index set $M_{o}$.  The subscript `$o$' refers to the interpretation of $M_{o}$ as corresponding to the `oracle' subset of covariates.  Moreover, $\beta_{M_{o}}^{0}$ denotes the true values of the oracle coefficients, while $\beta_{M_{o}}$ is understood as a random vector whose uncertainty resides in not knowing the true coefficients $\beta_{M_{o}}^{0}$.  For any subset $M$ the vector $\beta_{M}^{0}$ refers to the projection of the column space of $X_{M}$ on the true coefficients $\beta_{M_{o}}^{0}$, that is, $\beta_{M}^{0} = (X'_{M}X_{M})^{-1}X'_{M}X_{M_{o}}\beta_{M_{o}}^{0} = \E_{y}(\widehat{\beta}_{M})$.  Lastly, $\sigma_{M_{o}}^{0} > 0$ denotes the true unknown error standard deviation, while $\sigma_{M_{o}}$ is a random variable whose distribution expresses the uncertainty from not knowing $\sigma_{M_{o}}^{0}$, under the oracle model.  

The objective is to construct a statistical procedure which can be shown, asymptotically and demonstrated empirically, to be able to identify $M_{o}$ as the index set of the oracle model within the sample space of all $2^{p}$ candidate subsets $M \subset \{1,\dots,p\}$.  For each index set, $M$, in the sample space the conditional sampling distribution of the data is assumed as
\vspace{-.18cm}\begin{equation}\small\label{conditionalModel}
Y | \beta_{M}, \sigma_{M}^{2} \sim \text{N}_{n}\Big(X_{M}\beta_{M}, \sigma_{M}^{2}I_{n}\Big).
\vspace{-.18cm}\end{equation}
The centerpiece of our methodology is then the following definition.  The function $|\cdot|$ denotes the absolute value function if its argument is scalar-valued, and denotes the cardinality function if its argument is set-valued.  The norms $\|\cdot\|_{2}$ and $\|\cdot\|_{0}$ refer, respectfully, to the usual $L_{2}$ and $L_{0}$ norms defined on finite-dimensional Euclidean spaces.
\begin{definition}\label{admissibility}
Assume fixed $\varepsilon > 0$.  A given $\beta_{M}$ coupled within some index subset $M \subset \{1,\dots,p\}$ is said to be $\varepsilon$-$admissible$ if and only if $h(\beta_{M}) = 1$, where
\vspace{-.18cm}\begin{equation}\small\label{hFunction} 
h(\beta_{M}) := \I\Big\{\frac{1}{2}\|X'(X_{M}\beta_{M} - Xb_{\min})\|_{2}^{2} \ge \varepsilon \Big\},
\vspace{-.18cm}\end{equation}
and $b_{\min}$ solves $\ds \min_{b\in\R^{p}} \frac{1}{2}\|X'(X_{M}\beta_{M} - Xb)\|_{2}^{2}$ subject to $\|b\|_{0} \le |M|-1$.  
\end{definition}

Observe that this definition is consistent with the heuristic description of $\varepsilon$-$admissible$ subsets given in the previous section.  In particular, if the subset of covariates indexed by $M$ is linearly dependent or if one of the components of $\beta_{M}$ is zero, then $h(\beta_{M}) = 0$.  The subtlety in this definition is assuming an appropriately chosen $\varepsilon$ which is able to strike an optimal balance for distinguishing signal from noise.  Intuitively, $\varepsilon = \varepsilon(n,p,M)$, i.e., is a function of the amount of information available given by $n$, the difficulty of the problem represented by $p$, and information about a given $M$ being considered such as $|M|$.  For instance, if $|M| > n$ then $h(\beta_{M}) = 0$ because $X_{M}$ cannot have full rank.  In this case any $\varepsilon > 0$ will work, but the choice of $\varepsilon$ matters a lot if $|M| \le n$.  The choice of $\varepsilon$ is a major focus of Section \ref{TheoryResults} where the main results of the paper are presented, and from where we suggest the following default choice:
\vspace{-.18cm}\begin{equation}\small\label{epsilon_used}
\varepsilon = \Lambda_{M}\widehat{\sigma}_{M}^{2} \Big( \frac{n^{0.51}}{9} + |M|\frac{\log(p\pi)^{1.1}}{9} -  p_{o}\Big)_{+},
\vspace{-.18cm}\end{equation}
where $\Lambda_{M} := \tr\big((H_{M}X)'H_{M}X\big)$ and $\widehat{\sigma}_{M}^{2} := \RSS_{M} / (n - |M|)$ with $\RSS_{M} := y'(I_{n} - H_{M})y$ and $H_{M} := X_{M}(X'_{M}X_{M})^{-1}X'_{M}$, and the vector $y$ an observation from the true model (\ref{trueModel}).  The parameter $p_{o}$ represents prior belief about $|M_{o}|$, the number of covariates in the true model $M_{o}$.  In practice, a value of $p_{o}$ can be directly specified or selected by cross-validation.  A built-in cross-validation procedure is included in the accompanying software to this paper.  Details are provided with the simulation study in Section \ref{SimResults}.

Within the $h$ function in Definition \ref{admissibility} the quantity $\frac{1}{2}\|X'(X_{M}\beta_{M} - Xb_{\min})\|_{2}^{2}$ represents the difference in prediction for a subset $M$ against all subsets with fewer covariates.  This measure of distance has been adapted from \cite{Candes2007}, but they deal with the error $\|X'(y - Xb)\|_{\infty}$ over $b \in \R^{p}$.  This is very different from using $X_{M}\beta_{M}$ in place of $y$ because the former results in a noiseless measure of distance.  To illustrate, observe that
\vspace{-.18cm}\[\small
\E_{y}\big(\|X'(Y - Xb)\|_{2}^{2}\big) 
= \|X'(X_{M_{o}}\beta_{M_{o}} - Xb)\|_{2}^{2} +\sigma_{M_{o}}^{2} \cdot p, \\
\vspace{-.18cm}\]
where $\E_{y}(\cdot)$ is used to denote the expectation taken with respect to the sampling distribution of the data $Y$.

There are various reasons for using the quantity $X'(X_{M}\beta_{M} - Xb)$ from the Dantzig selector \citep{Candes2007} versus simply the difference in predictions $(X_{M}\beta_{M} - Xb)$ as in the LASSO \citep{Tibshirani}.  One reason is that $X'(X_{M}\beta_{M} - Xb)$ accounts for difference in predictions as well as correlations with the explanatory data, as discussed in \cite{Berk2008}.  If the difference in predictions is small but is highly correlated with the design matrix, then it is likely that the smaller subset of covariates is unable to account for the effect of one or more of the covariates in $M$.  Thus, using $X'(X_{M}\beta_{M} - Xb)$ instead of just the difference in predictions is a method of controlling for potential omitted variable effects which could incorrectly find a close fitting subset to $M$.  Another advantage of $X'(X_{M}\beta_{M} - Xb)$ is that it is invariant under orthogonal transformations of the design matrix, as pointed out in \cite{Candes2007}.

Now that the foundation for $\varepsilon$-$admissible$ subsets of the parameter space has been laid out, it remains to show how Definition \ref{admissibility} can be coupled with a likelihood based approach for constructing a probability distribution over index subsets $M \subset \{1,\dots,p\}$.  This is a common strategy for Bayesian approaches, i.e., construct a prior density with desirable properties for variable selection and then couple the prior with a likelihood function of the data to study the resulting posterior distribution.  However, it is not clear what sort of a prior to use within our $\varepsilon$-$admissible$ subsets approach, and recent developments in generalized fiducial inference (GFI) offer a systematic method of deriving objective Bayes-like posterior distributions.  To illustrate as in \cite{Hannig2016}, suppose that some data $Y = G(U,\theta)$ for some deterministic data generating equation $G(\cdot,\cdot)$, some parameters $\theta$, and some random component $U$ whose distribution is independent of $\theta$ and is completely known.  The generalized fiducial distribution of $\theta$ is then given by
\vspace{-.18cm}\[\small
r(\theta | y) = \frac{  f(y,\theta) J(y,\theta)  }{  \int_{\Theta} f(y,\theta') J(y,\theta') d\theta'  },
\vspace{-.18cm}\]
where $f$ is the likelihood function and
\vspace{-.18cm}\[\small
J(y,\theta) = D\bigg( \frac{d}{d\theta}G(u,\theta)\Big|_{u=G^{-1}(y,\theta)} \bigg)
\vspace{-.18cm}\]
with $D(A) = (\det A'A)^{\frac{1}{2}}$.  The component $J(y,\theta)$ is termed the $Jacobian$ because it results from inverting the data generating equation on the data.  We are committing a slight abuse of notation as $r(\theta|y)$ is not a conditional density in the usual sense.  Instead, we are using this notation to stress that the generalized fiducial distribution is a function of the observed data $y$.

To make matters concrete in the linear model setting of (\ref{conditionalModel}), the parameters are $\theta = (\beta_{M},\sigma_{M})$, the data generating equation is specified as $G\big(U, (\beta_{M},\sigma_{M})\big) = X_{M}\beta_{M} + \sigma_{M}U$ where $U \sim \text{N}_{n}(0,I_{n})$, and the Jacobian term reduces to $J\big(y, (\beta_{M},\sigma_{M})\big) = \sigma_{M}^{-1}|\det(X'_{M}X_{M})|^{\frac{1}{2}}\RSS_{M}^{\frac{1}{2}}$.  Thus, 
\vspace{-.18cm}\[\small
r\big((\beta_{M},\sigma_{M}) | y\big) \propto \sigma_{M}^{-n}e^{-\frac{\|y - X_{M}\beta_{M}\|^{2}_{2}}{2\sigma_{M}^{2}}}\sigma_{M}^{-1}|\det(X'_{M}X_{M})|^{\frac{1}{2}}\RSS_{M}^{\frac{1}{2}} \cdot h(\beta_{M}),
\vspace{-.18cm}\]
where the factor of $h(\beta_{M})$ appears in the likelihood from only considering $\varepsilon$-$admissible$ subsets of the parameter space.  Accordingly, as is done with a Bayesian posterior density and in Section~3 of \cite{Hannig2016}, define the GFI probability of a given subset $M$ to be proportional to the normalizing constant of $r\big((\beta_{M},\sigma_{M}) | y\big)$.  That is,
\vspace{-.18cm}\[\small
\begin{split}
r(M|y) & := \frac{  \int f\big(y,(\beta_{M},\sigma_{M})\big) J\big(y,(\beta_{M},\sigma_{M})\big) h(\beta_{M}) \ d(\sigma_{M},\beta_{M})  }{  \sum\limits_{j=1}^{p}\sum\limits_{|M|=j}\int f\big(y,(\beta_{M},\sigma_{M})\big) J\big(y,(\beta_{M},\sigma_{M})\big) h(\beta_{M}) \ d(\sigma_{M},\beta_{M})  } \\
& \propto \int_{\R^{p_{M}}}\int_{0}^{\infty}h(\beta_{M})\frac{|\det(X'_{M}X_{M})|^{\frac{1}{2}}\RSS_{M}^{\frac{1}{2}}}{\sigma_{M}^{n+1}e^{\frac{(y-X_{M}\beta_{M})'(y-X_{M}\beta_{M})}{2\sigma_{M}^{2}}}} \ d\sigma_{M} \ d\beta_{M}, \\
\end{split}
\vspace{-.18cm}\]
which simplifies to 
\vspace{-.18cm}\begin{equation}\small\label{ModelPMF}
r(M|y) \propto \pi^{\frac{|M|}{2}}\Gamma\Big(\frac{n-|M|}{2}\Big)\RSS_{M}^{-(\frac{n-|M|-1}{2})}\E(h(\beta_{M})),
\vspace{-.18cm}\end{equation}
where the expectation is taken with respect to the location-scale multivariate T distribution,
\vspace{-.18cm}\begin{equation}\small\label{tDist}
t_{n-|M|}\Big(\widehat{\beta}_{M},\frac{\RSS_{M}}{n-|M|}(X_{M}'X_{M})^{-1}\Big)
\vspace{-.18cm}\end{equation}
with $\widehat{\beta}_{M} := (X_{M}'X_{M})^{-1}X_{M}'y$. 
Notice that the quantity $\E(h(\beta_{M}))$ is a function of the observed data $y$.

Observe that (\ref{ModelPMF}) expresses the relative likelihood of the subset $M$ over all $2^{p}$ possible subsets.  The expression can be described as a product of two terms, the first being comprised of information from the sampling distribution of the data and largely driven by the residual sum of squares, $\RSS_{M}$, and the second having to do with the $\varepsilon$-$admissibility$ of $\beta_{M}$, in the form of $\E(h(\beta_{M}))$.  Thus, the support of $r(M|y)$ in (\ref{ModelPMF}) is dominated by the $\varepsilon$-$admissible$ subsets, as desired.  

Section \ref{TheoryResults} provides the conditions and supporting lemmas and theorems needed to show that $r(M_{o}|Y) \to 1$ in probability as $n, p \to \infty$.  First however, a few remarks are provided about computing $r(M|y)$ on actual data.

\subsection{Remarks on computation}\label{computationRemarks}

With a probability distribution now defined over $\varepsilon$-$admissible$ subsets, it must be demonstrated that $r(M|y)$ in (\ref{ModelPMF}) can be efficiently computed.  There are two main computational issues to deal with.  The first is to evaluate $h(\beta_{M})$ for a given $\beta_{M}$, and the second is to sample subsets $M$ via pseudo-marginal based MCMC.  The computational complexity and the need for pseudo-marginal based MCMC arises because neither $h(\beta_{M})$ nor $\E(h(\beta_{M}))$ have a closed form solution.  

To evaluate $h(\beta_{M})$ for a given $\beta_{M}$ we adapt an explicit $L_{0}$ minimization algorithm introduced in \cite{Bertsimas2016}.  The authors state that their algorithm borrows ideas from projected gradient descent and methods in first-order convex optimization, and solves problems of the form $\min_{b\in\R^{p}} g(b)$ subject to $\|b\|_{0} \le \kappa$, where $g(b) \ge 0$ is convex and has Lipschitz continuous gradient: $\|\nabla g(b) - \nabla g(\widetilde{b})\|_{2} \le l\|b - \widetilde{b}\|_{2}$.  The algorithm is not guaranteed to find a global optimum (unless formal optimality tests are run, which can take a long time), but \cite{Bertsimas2016} provide provable guarantees that the algorithm will converge to a first-order stationary point, which is defined as a vector $\tilde{b} \in \R^{p}$ with $\|\tilde{b}\|_{0} \le \kappa$ which satisfies $\tilde{b} = \tilde{b} - \frac{1}{l}\nabla g(\tilde{b})$.  Paraphrasing from \cite{Bertsimas2016}, their algorithm detects the active set after a few iterations, and then takes additional time to estimate the coefficient values to a high accuracy level.  In our application of their algorithm we are $not$ first-most interested in finding a global optimum.  To evaluate $h(\beta_{M})$, we need only determine if $\min_{b\in\R^{p}} \frac{1}{2}\|X'(X_{M}\beta_{M} - Xb)\|_{2}^{2}$ is smaller than $\varepsilon$ (as in (\ref{epsilon_used})).  For $\beta_{M}$ which are not $\varepsilon$-$admissible$, the objective function, $\frac{1}{2}\|X'(X_{M}\beta_{M} - Xb)\|_{2}^{2}$, can be made small very quickly via our implementation of the $L_{0}$ minimization algorithm.  To illustrate how consider the following specifics of our implementation.  The precise details regarding the algorithm can be found accompanying our software documentation at \verb1https://jonathanpw.github.io/software.html1.

First, to estimate $\E(h(\beta_{M}))$ we use a sample mean of sample vectors drawn from the location-scale multivariate T distribution in (\ref{tDist}).  This multivariate T distribution is centered at the least squares estimator, $\widehat{\beta}_{M}$, and multivariate theory suggests that $\widehat{\beta}_{M}$ will on average be close to the coefficients $\beta_{M}^{0}$.  By warm starting the $L_{0}$ minimization algorithm at $\widehat{\beta}_{M}$ with the smallest coefficient removed, subsets corresponding to $\beta_{M}^{0}$ with at least one zero coefficient typically yield $h(\beta_{M}) = 0$ within a few steps of the algorithm.

Second, as per the definition of $h(\cdot)$ in (\ref{hFunction}) the objective function is minimized over all $b \in \R^{p}$ with $\|b\|_{0} \le |M| - 1$.  Hence, the $\kappa$ required for the $L_{0}$ minimization algorithm from \cite{Bertsimas2016} is naturally chosen for us as $\kappa = |M|-1$.  Knowing how to choose $\kappa$ greatly reduces the $L_{0}$ optimization problem.  Moreover, our implementation is further simplified by the fact that the closest prediction to $X_{M}\beta_{M}$ for a given $M$ is guaranteed to have $|M| - 1$ covariates.  Accordingly, the objective function in $h(\beta_{M})$ need $not$ be minimized over all $b \in \R^{p}$ with $\|b\|_{0} \le |M| - 1$, but can be minimized over all $b \in \R^{p}$ with $\|b\|_{0} = |M| - 1$.

The second computational issue is to sample subsets $M$ via pseudo-marginal based MCMC.  We do this by using the Grouped Independence Metropolis Hastings (GIMH) algorithm from \cite{Andrieu2009}, but originally introduced in \cite{Beaumont2003}.  The reason standard MCMC techniques do not apply is that there is no obvious closed form expression for the probability mass function (\ref{ModelPMF}) because of the expectation, $\E(h(\beta_{M}))$, in the expression.  As described in \cite{Andrieu2009} such situations warrant introducing a latent variable to yield analytical expressions or easier implementation.  

In the case of $r(M|y)$ in (\ref{ModelPMF}), we introduce the latent location-scale multivariate T vector in (\ref{tDist}) from within the expectation $\E(h(\beta_{M}))$.  Our pseudo-marginal based MCMC is carried out by sampling an index subset $M$ along with sampling some pre-specified number, $N$, of multivariate T vectors (corresponding to $M$) from (\ref{tDist}).  The sample of multivariate T vectors, say $B$, is then used to compute the sample mean estimate of $\E(h(\beta_{M}))$.  Accordingly, we define a joint Markov chain on $(M,B)$, but discard $B$ to obtain samples from the marginal distribution of $M$.  As argued in \cite{Andrieu2009}, this is a valid MCMC sampling strategy, but is known to suffer from slower mixing than if we were able to integrate the $\beta_{M}$ out of the mass function $r(M|y)$ in (\ref{ModelPMF}), i.e., analytically evaluate $\E(h(\beta_{M}))$.  However, this is not possible given the $h$ function in (\ref{hFunction}).  Additionally, the mixing associated with pseudo-marginal approaches is known to be poor when the number of importance samples ($N$, the sample size of $B$) is small.  These practical bottlenecks outline avenues for future research.  Nonetheless, we demonstrate in Section \ref{SimResults} that our computational strategies are efficient enough to be implemented on actual data, in comparison to other common penalized likelihood and Bayesian approaches.

\section{Theoretical results}\label{TheoryResults}

The main objective of this section is to show under what conditions, asymptotically, $r(M_{o}|Y)$ in (\ref{ModelPMF}) will converge to 1, particularly if $p >> n$.  The $\varepsilon$-$admissible$ subsets approach is able to achieve such a strong consistency result because the resulting sample space is effectively reduced to only those subsets with no redundancies.  The essence of the mathematical result is that the space of $\varepsilon$-$admissible$ sets is small enough that the true model can be detected.  This addresses the issue raised in \cite{Ghosh2015} that high-dimensional settings often lead to arbitrarily small probabilities for all models (including the true model) simply because there are too many models to consider.

\subsection{Discussion of the conditions}

The first two conditions, Condition \ref{condition1} and Condition \ref{condition2}, are to ensure that the true model, $M_{o}$, is identifiable.  Observe from (\ref{hFunction}) that $\varepsilon$ is used to control the sensitivity and specificity for identifying $\varepsilon$-$admissible$ subsets.  In particular, if $\varepsilon$ is too large, then $h(\beta_{M_{o}})$ will incorrectly be set to zero implying that $\beta_{M_{o}}$ is not $\varepsilon$-$admissible$.  Condition \ref{condition1} specifies how large $\varepsilon$ can be whilst the true model remains identifiable.  This condition turns out to be critically important in actual data applications because computing $h(\beta_{M_{o}})$ is closely related to the comparison in equation (\ref{trueModelComp}).

\begin{condition}\label{condition1}
For large $n$ and $p$,
\vspace{-.18cm}\begin{equation}\small\label{trueModelComp}
\frac{1}{18}\|X'(X_{M_{o}}\beta_{M_{o}}^{0} - Xb_{\min})\|_{2}^{2} \ge \varepsilon
\vspace{-.18cm}\end{equation}
where $b_{\min}$ solves $\ds \min_{b\in\R^{p}} \frac{1}{2}\|X'(X_{M_{o}}\beta_{M_{o}}^{0} - Xb)\|_{2}^{2}$ subject to $\|b\|_{0} \le |M_{o}|-1$.
\end{condition}

Condition \ref{condition2} is born from Lemma \ref{RSSLemma} which is an important necessary result for the main result of this paper, Theorem \ref{MainResult}.  The term $\log(n)^{\gamma}$ represents the sparsity assumption for the true model, i.e., the number of covariates in the true model must not exceed $\log(n)^{\gamma}$ for some fixed scalar $\gamma > 0$.  The $\gamma$ parameter indicates that the asymptotic results remain true if the true model grows faster than $\log(n)$, but not faster than some power of $\log(n)$.  In finite-sample applications, $\gamma$ has no consequence and can be ignored.

The constant $\alpha \in (0,1)$ reflects the only explicit restriction needed on the sample space of $2^{p}$ subsets to show that $r(M_{o}|Y) \to 1$ in probability for large $n$ and $p$, Theorem \ref{MainResult}.  The residual sum of squares term in $r(M|Y)$ in (\ref{ModelPMF}) cannot be controlled (as a ratio to $r(M_{o}|Y)$) for arbitrary subsets $M$ with $|M| = O(n)$ because the column span of $X_{M}$ includes $y \in \R^{n}$ when rank$(X_{M}) = n$.  To eliminate such subsets from the sample space, Condition \ref{condition2} requires that only subsets of size $|M| \le n^{\alpha}$ can be given nonzero probability.  However, recall from Definition \ref{admissibility} that $h(\beta_{M}) = 0$ if $|M| > n$ because in this case the columns of $X_{M}$ must be linearly dependent.  Accordingly, all subsets $M$ with $|M| > n$ are given zero probability, by definition.  Evidenced by this fact, the only explicit restriction placed on the sample space is that subsets $M$ with $|M| \in (n^{\alpha},n)$ are excluded.  In sparse settings it is assumed that $|M_{o}| << n$ anyway, so neglecting such subsets is reasonable.  Convergence to the true model $M_{o}$ will be quicker for smaller $\alpha$ because there are less models to consider, but too small of an $\alpha$ will exclude $M_{o}$ from the sample space.

In Condition \ref{condition2}, and for the remainder of this section assume that $\gamma > 0$, say $\gamma = 1$, and $\alpha \in (0,1)$, say $\alpha = .5$, have been chosen and fixed at appropriate values.

\begin{condition}\label{condition2}
The true model $M_{o}$ satisfies $|M_{o}| \le \log(n)^{\gamma}$, and 
\vspace{-.18cm}\[\small
\lim_{\substack{n\to\infty\\p\to\infty}}\min\Big\{\frac{\Delta_{M}}{|M_{o}|\log(p)} : M \ne M_{o}, |M| \le |M_{o}|\Big\} = \infty,  
\vspace{-.18cm}\]
\vspace{-.18cm}\[\small
\liminf_{\substack{n\to\infty\\p\to\infty}}\frac{n^{1-\alpha}}{\log(p)} > 2, \ \ \text{and} \ \ \log(p) < \frac{n-|M_{o}|-1}{4\log(n)^{\gamma}},
\vspace{-.18cm}\]
for large $n$ and $p$, where $\Delta_{M} := \|X_{M_{o}}\beta_{M_{o}}^{0} - H_{M}X_{M_{o}}\beta_{M_{o}}^{0}\|_{2}^{2}$ as in \cite{Lai2015}.
\end{condition}
This is a slightly weaker version of condition (11) in \cite{Lai2015}.  They relate Condition \ref{condition2} to the sparse Riesz condition \citep{Zhang2008} which requires that the eigenvalues of $X_{M}'X_{M}/n$ are uniformly bounded away from 0 and $\infty$.  Essentially, $\Delta_{M}$ is a measure of how distinct the true model predictions $X_{M_{o}}\beta_{M_{o}}^{0}$ are from their projection onto the column space of $X_{M}$ for $M \ne M_{o}$ and $|M| \le |M_{o}|$.  Recall that $H_{M} := X_{M}(X'_{M}X_{M})^{-1}X'_{M}$.  In particular, if $X_{M}$ is orthogonal to $X_{M_{o}}$, then $\Delta_{M} = \|X_{M_{o}}\beta_{M_{o}}^{0}\|_{2}^{2} $ which will be much larger than the denominator, $|M_{o}|\log(p)$.  The requirements of this condition are reasonable because $\Delta_{M}$ grows very fast for $M$ such that $M_{o} \not\subseteq M$. 

Condition \ref{condition2} is important for being able to identify the true model amongst other models $M$ with $|M| \le |M_{o}|$.  The next two conditions address the requirements for $M$ with $|M| > |M_{o}|$, which primarily rely on the fact that such subsets are not $\varepsilon$-$admissible$.

Conditions \ref{condition3} and \ref{condition4} demonstrate how large $\varepsilon$ needs to be to achieve the consistency result of the main theorem.  Condition \ref{condition3} states that for subsets of covariates with redundancies, $\varepsilon$ needs to be larger than the difference in projections of the true model prediction, $X_{M_{o}}\beta_{M_{o}}^{0}$, onto $M$ and onto a strict subset of $M$.  This condition facilitates the intuition that the variable selection procedure will not concentrate on subsets $M$ with redundant covariates.  If a given subset $M$ is not $\varepsilon$-$admissible$, then the difference in projections will be small so that the condition is easily achieved.

\begin{condition}\label{condition3}
For any $M$ with $|M| > |M_{o}|$, for large $n$ or $p$,
\vspace{-.18cm}\[\small
\frac{9}{2}\|X'(H_{M} - H_{M(-1)})X_{M_{o}}\beta_{M_{o}}^{0}\|_{2}^{2} < \varepsilon,
\vspace{-.18cm}\]
where $H_{M(-1)}$ is the projection matrix for $M$ after omitting the covariate which minimizes $\|X'(H_{M} - H_{M(-1)})X_{M_{o}}\beta_{M_{o}}^{0}\|_{2}^{2}$.
\end{condition}
In fact, if $M_{o} \subset M$ with $|M_{o}| < |M|$, then $H_{M}X_{M_{o}}\beta_{M_{o}}^{0} = H_{M(-1)}X_{M_{o}}\beta_{M_{o}}^{0}$ in which case Condition \ref{condition3} holds trivially.  

Lastly, Condition \ref{condition4} describes the rate at which $\varepsilon$ needs to grow to achieve the consistency of the main result.  The distinction between Condition \ref{condition3} and Condition \ref{condition4} is that the former provides a necessary lower bound for arguing that $\E(h(\beta_{M}))$ vanishes for $M$ such that $|M| > |M_{o}|$, while the latter provides the rate at which $\varepsilon$ must grow to achieve the consistency result of Theorem \ref{MainResult}.  The terms which compete with $\varepsilon$ arise in the proofs of Lemma \ref{RSSLemma} and Theorem \ref{MainResult}.  Recall that $\widehat{\sigma}_{M}^{2} := \RSS_{M} / (n - |M|)$, where $\RSS_{M}$ is the classical residual sum of squares for model $M$, and that $\Lambda_{M} := \tr\big((H_{M}X)'H_{M}X\big)$.

\begin{condition}\label{condition4}
\vspace{-.18cm}\[\small
\lim_{\substack{n\to\infty\\p\to\infty}}\min_{|M|\le n^{\alpha}} \frac{ \frac{\varepsilon}{18\Lambda_{M}\widehat{\sigma}_{M}^{2}} + D_{1}|M_{o}| - \varphi(M,n,p) }{\log(n)} = \infty,
\vspace{-.18cm}\]
where
\vspace{-.18cm}\[\small
\varphi(M,n,p) := 4e^{2} n^{\alpha} + (D_{1}+(1 + 4e^{2})\log(p))|M| + \log\Big(\frac{|M|}{1 - \frac{1}{n-|M|}}\Big),
\vspace{-.18cm}\]
and $D_{1} = \frac{1}{2}\log\Big(\frac{6\pi}{1-\frac{n^{\alpha}+2}{n}}\Big)$.  Additionally, $\frac{\varepsilon}{9\Lambda_{M}\widehat{\sigma}_{M}^{2}} < \frac{n-|M|}{2}$ for all $M$ with $|M|\le n^{\alpha}$.
\end{condition}
The $\Lambda_{M}$ term arises from Lemmas \ref{Term1Bound} and \ref{Term2Bound}.  It is intimately related to the presence of collinearity amongst the covariates, and Condition \ref{condition4} implies that $\varepsilon$ must account for collinearity by controlling for $\Lambda_{M}$.  Observe that if $X$ is orthogonal, then $\Lambda_{M} = |M|$.

\subsection{Main result}

The first two results are lemmas which are needed in the proofs of Theorems \ref{BigModel} and \ref{TrueModelCase}.  Lemma \ref{Term1Bound} illustrates the rate at which $\beta_{M}$ concentrates around its mean, $\widehat{\beta}_{M}$, the least squares estimator, and Lemma \ref{Term2Bound} illustrates the rate at which $\widehat{\beta}_{M}$ concentrates around its mean, $\E_{y}(\widehat{\beta}_{M})$.  Theorem \ref{BigModel} uses these two lemmas to bound the rate at which $\beta_{M}$ concentrates around $\E_{y}(\widehat{\beta}_{M})$ for subsets $M$ with $|M| > |M_{o}|$.  This yields an upper bound on $\E(h(\beta_{M}))$ with a probabilistic guarantee, and implies that $\E(h(\beta_{M}))$ vanishes for large $n$ and $p$, for large non-$\varepsilon$-$admissible$ subsets.  The proofs are relegated to Appendix \ref{AppendProofs}.

\begin{lemma}\label{Term1Bound}
For any fixed $c_{1} \in (0,1)$ assume $|M| \le c_{1}n$, and choose $n$ and $p$ such that $\frac{\varepsilon}{9\Lambda_{M}\widehat{\sigma}_{M}^{2}} < \frac{n-|M|}{2}$.  If
\vspace{-.18cm}\[\small
\beta_{M} \sim \text{t}_{n-|M|}\Big(\widehat{\beta}_{M}, \frac{\RSS_{M}}{n-|M|}(X_{M}'X_{M})^{-1}\Big),
\vspace{-.18cm}\]
where $\widehat{\beta}_{M} = (X_{M}'X_{M})^{-1}X_{M}'y$, then
\vspace{-.18cm}\[\small
P\Big(\frac{1}{2}\|X'X_{M}(\beta_{M} - \widehat{\beta}_{M})\|_{2}^{2} \ge \frac{\varepsilon}{9}\Big) \le \frac{2^{\frac{3}{2}}3|M|\widehat{\sigma}_{M}\sqrt{\Lambda_{M}}e^{-\frac{\varepsilon}{18\Lambda_{M}\widehat{\sigma}_{M}^{2}}}}{\sqrt{\pi\varepsilon}(1 - \frac{1}{n-|M|})}.
\vspace{-.18cm}\]
\end{lemma}

In the next lemma, $P_{y}$ is used to denote the probability measure associated with the sampling distribution of the data $Y$.

\begin{lemma}\label{Term2Bound}
Assume $|M| < n$, and $Y | \beta_{M}, \sigma_{M}^{2} \sim \text{N}_{n}\Big(X_{M}\beta_{M}, \sigma_{M}^{2}I_{n}\Big)$.  Then the classical least squares estimator $\widehat{\beta}_{M} \sim \text{N}(\E_{y}(\widehat{\beta}_{M}), \sigma^{2}_{M}(X_{M}'X_{M})^{-1})$, and 
\vspace{-.18cm}\[\small
P_{y}\Big(\frac{1}{2}\|X'X_{M}\big(\widehat{\beta}_{M} - \E_{y}(\widehat{\beta}_{M})\big)\|_{2}^{2} \ge \frac{\varepsilon}{9}\Big) \le \frac{3|M|\sigma_{M}\sqrt{\Lambda_{M}}}{\sqrt{\pi\varepsilon}}e^{-\frac{\varepsilon}{9\sigma^{2}_{M}\Lambda_{M}}}.
\vspace{-.18cm}\]  
\end{lemma} 

Combining these two lemmas gives the following non-asymptotic concentration result for models that are larger than the true model.  Recall that the expectation $\E(h(\beta_{M}))$ depends on the observed data $y$. The following two theorems study the frequentist behavior of this quantity with respect to the sampling distribution of $Y$.

\begin{theorem}\label{BigModel}
For any fixed $c_{1} \in (0,1)$ suppose $|M_{o}| < |M| \le c_{1}n$, choose $n$ and $p$ such that $\frac{\varepsilon}{9\Lambda_{M}\widehat{\sigma}_{M}^{2}} < \frac{n-|M|}{2}$, and assume that $\varepsilon$ satisfies Condition \ref{condition3}.  Then
\vspace{-.18cm}\[\small
P_{y}\bigg(\E(h(\beta_{M})) \le \frac{2^{\frac{3}{2}}3|M|\widehat{\sigma}_{M}\sqrt{\Lambda_{M}}}{\sqrt{\pi\varepsilon}(1 - \frac{1}{n-|M|})}e^{-\frac{\varepsilon}{18\Lambda_{M}\widehat{\sigma}_{M}^{2}}}\bigg) 
\ge 1 - \frac{3|M|\sigma_{M}\sqrt{\Lambda_{M}}}{\sqrt{\varepsilon\pi}e^{\frac{\varepsilon}{9\sigma^{2}_{M}\Lambda_{M}}}}.
\vspace{-.18cm}\]
\end{theorem}

The next result is a probabilistic guarantee the true model is $\varepsilon$-$admissible$ given $\varepsilon$ satisfies Condition \ref{condition1}.  This result is a statement that $M_{o}$ is identifiable.

\begin{theorem}\label{TrueModelCase}
For any fixed $c_{1} \in (0,1)$ suppose $|M_{o}| \le c_{1}n$, choose $n$ and $p$ such that $\frac{\varepsilon}{9\Lambda_{M_{o}}\widehat{\sigma}_{M_{o}}^{2}} < \frac{n-|M_{o}|}{2}$, and assume that $\varepsilon$ satisfies Condition \ref{condition1}.  Then 
\vspace{-.18cm}\[\small
P_{y}\Bigg(\E(h(\beta_{M_{o}})) \ge 1 - \frac{2^{\frac{3}{2}}3p_{M_{o}}\widehat{\sigma}_{M_{o}}\sqrt{\Lambda_{M_{o}}}}{\sqrt{\pi\varepsilon}(1 - \frac{1}{n-p_{M_{o}}})e^{\frac{\varepsilon}{18\Lambda_{M_{o}}\widehat{\sigma}_{M_{o}}^{2}}}}\Bigg)
\ge 1 - \frac{3p_{M_{o}}\sigma_{M_{o}}\sqrt{\Lambda_{M_{o}}}}{\sqrt{\varepsilon\pi}e^{\frac{\varepsilon}{9\sigma_{M_{o}}^{2}\Lambda_{M_{o}}}}}.
\vspace{-.18cm}\]
\end{theorem}

The following result is the main result of the paper.  It shows that the ratio of the generalized fiducial probability of the true model to the sum over that of all other subsets of covariates $M$ satisfying $|M| \le n^{\alpha}$ will converge to 1 in probability for large $n$ and $p$.  Note that the restriction to subsets $M$ with $|M| \le n^{\alpha}$ is a stronger restriction than $|M| \le c_{1}n$, which is sufficient for Theorems \ref{BigModel} and \ref{TrueModelCase}.  The reason being that the main result is stronger than the results of these two theorems.  In fact, Theorems \ref{BigModel} and \ref{TrueModelCase} are non-asymptotic results that hold for each fixed model $M$ separately, while Theorem \ref{MainResult} is an asymptotic result which applies uniformly over all $|M| \le n^{\alpha}$.  Just like with a conditional distribution, $r(M|Y)$ is obtained by replacing the observed data $y$ with the random variable $Y$ (random with respect to the sampling distribution), in \eqref{ModelPMF}.

\begin{theorem}\label{MainResult}
Given Conditions \ref{condition1}-\ref{condition4}, the true model $M_{o}$ satisfies,
\vspace{-.18cm}\[\small
\frac{r(M_{o}|Y)}{\sum_{j=1}^{n^{\alpha}}\sum_{M:|M|=j}r(M | Y)} \overset{P_y}{\longrightarrow} 1
\vspace{-.18cm}\]
as $n \to \infty$ or $n, p \to \infty$.
\end{theorem}

Although this is an asymptotic result, many of the ingredients that are used in its proof are non-asymptotic concentration results which are valid if Conditions \ref{condition1}-\ref{condition4} are satisfied.  Therefore, it can be expected that when these conditions are satisfied, in finite-sample situations the generalized fiducial distribution will concentrate on the true model $M_{o}$.  This expectation is indeed validated by the empirical performance of the procedure, which is demonstrated in the following section.

\section{Simulation results}\label{SimResults}

This section serves to demonstrate the empirical performance of our algorithm on synthetic data.  It is comprised of essentially two simulation setups.  The first setup, similar to that presented in \cite{Johnson2012}, compares our procedure to the nonlocal prior approach, the spike and slab LASSO of \cite{Rockova2015}, the elastic net as implementated in the $Python$ module \verb1scikit-learn1 \citep{scikit-learn}, and to the SCAD as implementated in the $R$ package \verb1ncvreg1 \citep{ncvreg}.  The authors of the nonlocal prior and the spike and slab LASSO, respectively, have made available the $R$ packages \verb1mombf1 and \verb1SSL1 for implementing their methods.  

The second setup illustrates a critical difference between our $\varepsilon$-$admissible$ subsets approach and the nonlocal prior approach.  Namely, for highly collinear finite-sample settings in which the true model is not uniquely expressed, given the level of noise in the data (i.e., $\sigma_{M_{o}}^{0}$), we demonstrate that our approach concentrates (in the sense of the MAP estimator) on subsets with fewer covariates without sacrificing prediction error.  The intuition for why this should be the case was discussed in Section \ref{Intro}.

\subsection{Simulation setup 1}

Here we generate 2000 data vectors $y$ according to model (\ref{trueModel}) with $M_{o}$ consisting of 8 covariates corresponding to $\beta_{M_{o}}^{0} = (-1.5, -1, -.8, -.6, .6, .8, 1, 1.5)'$, and $\sigma_{M_{o}}^{0} = 1$.  The $n\times p$ design matrix $X$ is generated with rows from the N$_{p}(0,\Sigma)$ distribution, where the diagonal components $\Sigma_{ii} = 1$ and the off-diagonal components $\Sigma_{ij} = \rho$ for $i \ne j$.  The first 1000 $y$ correspond to an independent design with $\rho = 0$, while the last 1000 $y$ correspond to $\rho = .25$, as in the simulation setup of \cite{Johnson2012}.  Note that 2000 design matrices $X$ are generated, and one $y$ is generated from each design.  The sample size $n$ is set at $n = 100$, and $p = 100, 200, 300, 400, 500$ are all considered.

We implement our algorithm on each of the 2000 synthetic data sets for 15000 MCMC steps with the first 5000 discarded.  Squared coefficient estimates from elastic net (using \verb1scikit-learn1) added by $n^{-2}$ serve as MCMC covariate proposal weights.  The default $\varepsilon$ in (\ref{epsilon_used}) is used, and we implemented a 10-fold cross-validation scheme for choosing our tuning parameter $p_{o}$ (prior to starting the algorithm).  The cross-validation consists of breaking the data into 10 folds (with a different set of 10 observations held out at each fold since $n = 100$), and implementing our MCMC algorithm separately for each $p_{o}$ in the grid $\{1,2,\dots,10\}$, on each of the 10 training sets.  Each of the 10 implementations of the MCMC on each of the 10 training sets is run for 200 steps with the first 100 steps discarded ($N = 30$ is set during the cross-validation procedure).  Squared nonzero coefficient estimates from elastic net (using \verb1scikit-learn1) serve as MCMC covariate proposal weights within the cross-validation procedure.  The MAP estimated subset for each MCMC chain is used to compute the Bayesian information criterion (BIC) on the held-out test set, and the computed BIC values are averaged over the 10 test sets, for each $p_{o} \in \{1,2,\dots,10\}$.  The $p_{o}$ corresponding to the minimum average test set BIC is then selected. 

Finally, for our implementation of the algorithm post-selection of $p_{o}$, the number of importance samples for estimating $\E(h(\beta_{M}))$ within each step of the algorithm is set at $N = 100$ which, through empirical experimentation, seems to be enough.  All competing variable selection procedures are implemented using existing software at default specifications.  The one exception is that the nonlocal prior procedure is set to run for 5000 steps, as is the case in the simulation setup of \cite{Johnson2012}.  The nonlocal prior procedure/software did not scale well for increased $p$, and required over a weeks worth of parallel computations on a computing cluster to obtain the results for the first simulation setup.  The tuning parameters for all methods are chosen with the default cross-validation procedures provided with the software.  Lastly, as in the simulation section for \cite{Rockova2015} their $\lambda_{1}$ is set at 1 (with a grid of 20 $\lambda_{0}$ values ending at 50), and their adaptive (best performing) procedure is used with $\theta \sim \text{Beta}(1,p)$.

\begin{figure}[H]
\centering
\includegraphics[trim=0cm 0cm 0cm 0cm, clip=true, scale=.38]{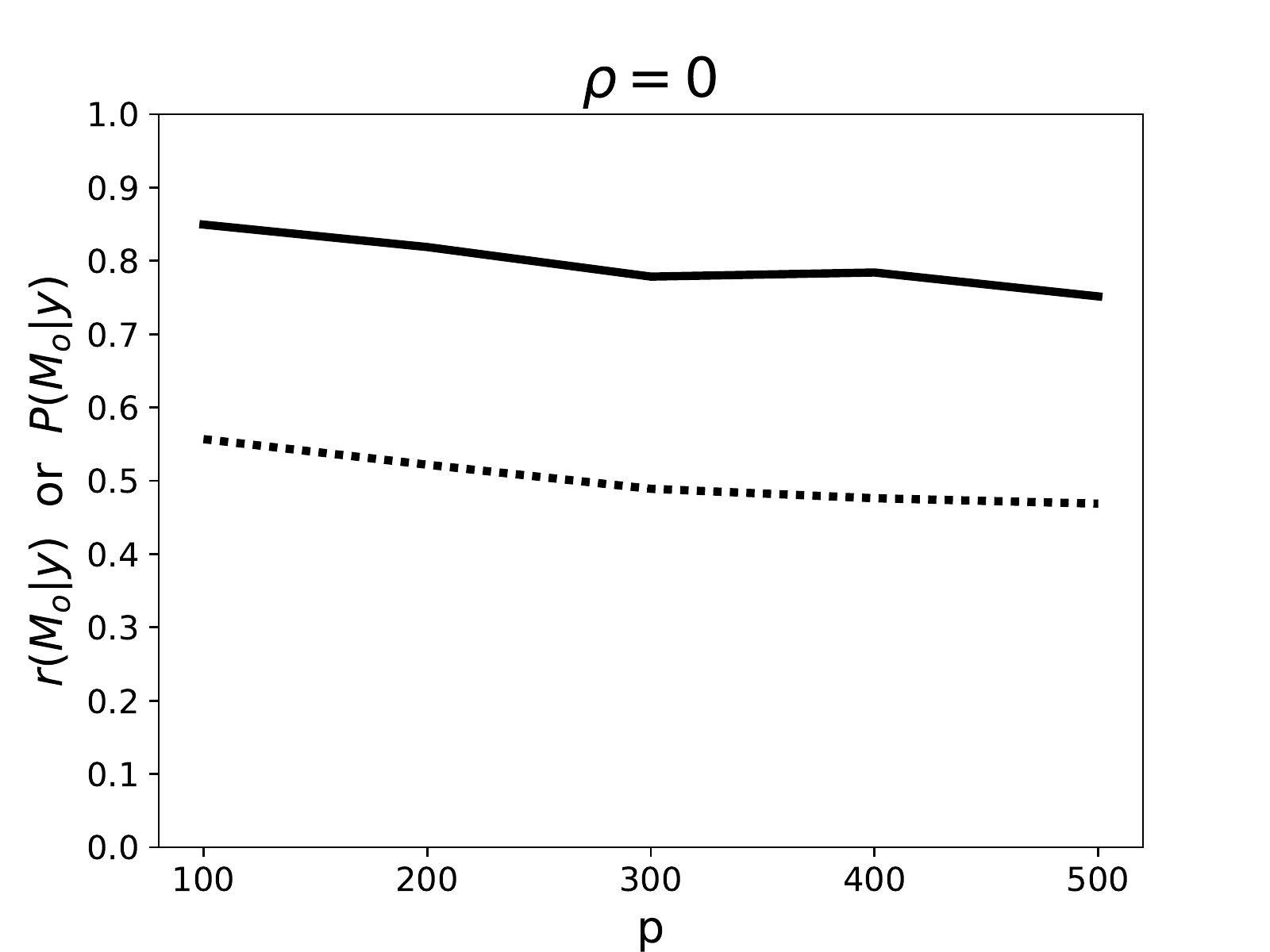}\includegraphics[trim=0cm 0cm 0cm 0cm, clip=true, scale=.38]{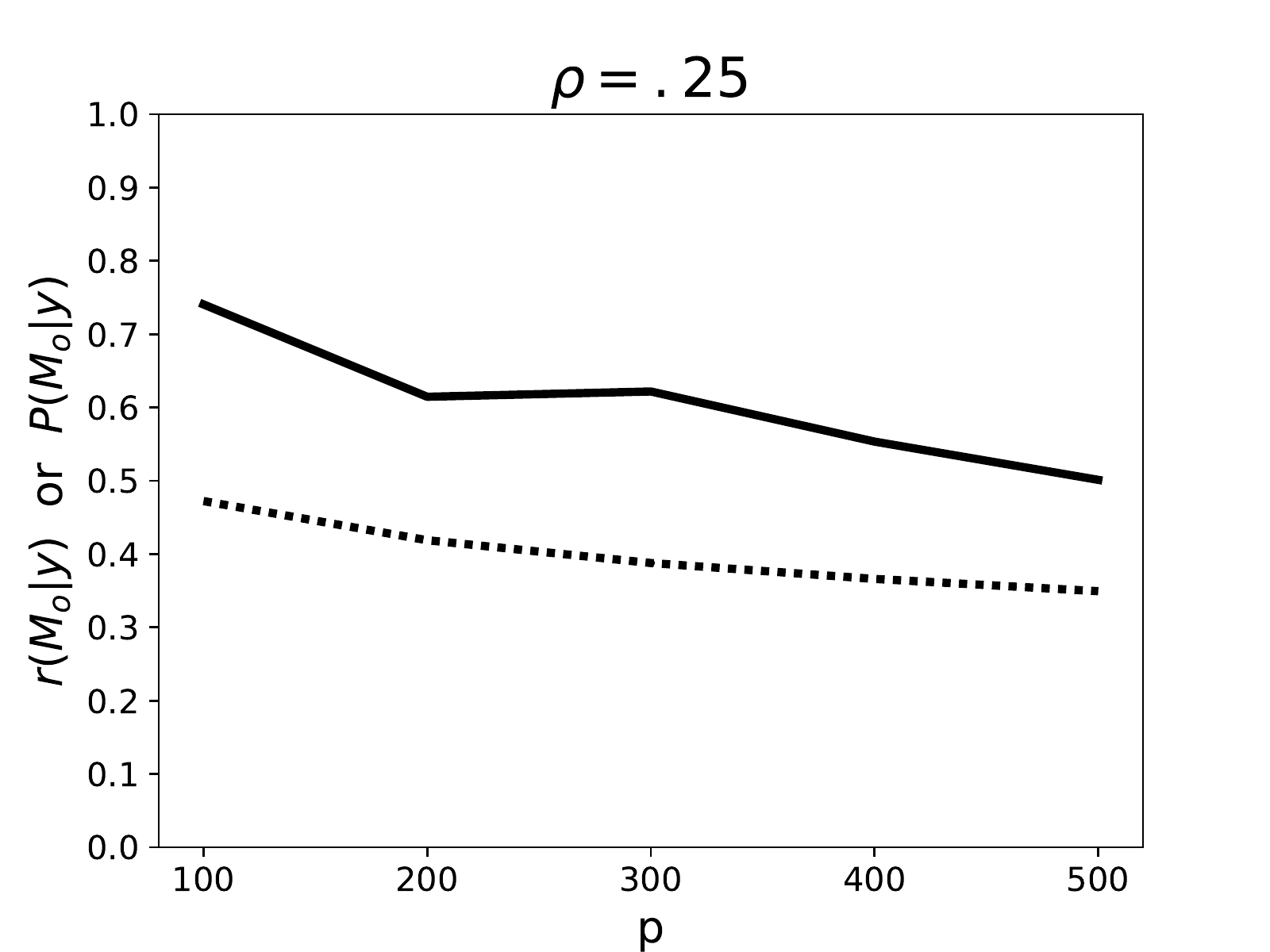}
\includegraphics[trim=0cm 0cm 0cm 0cm, clip=true, scale=.38]{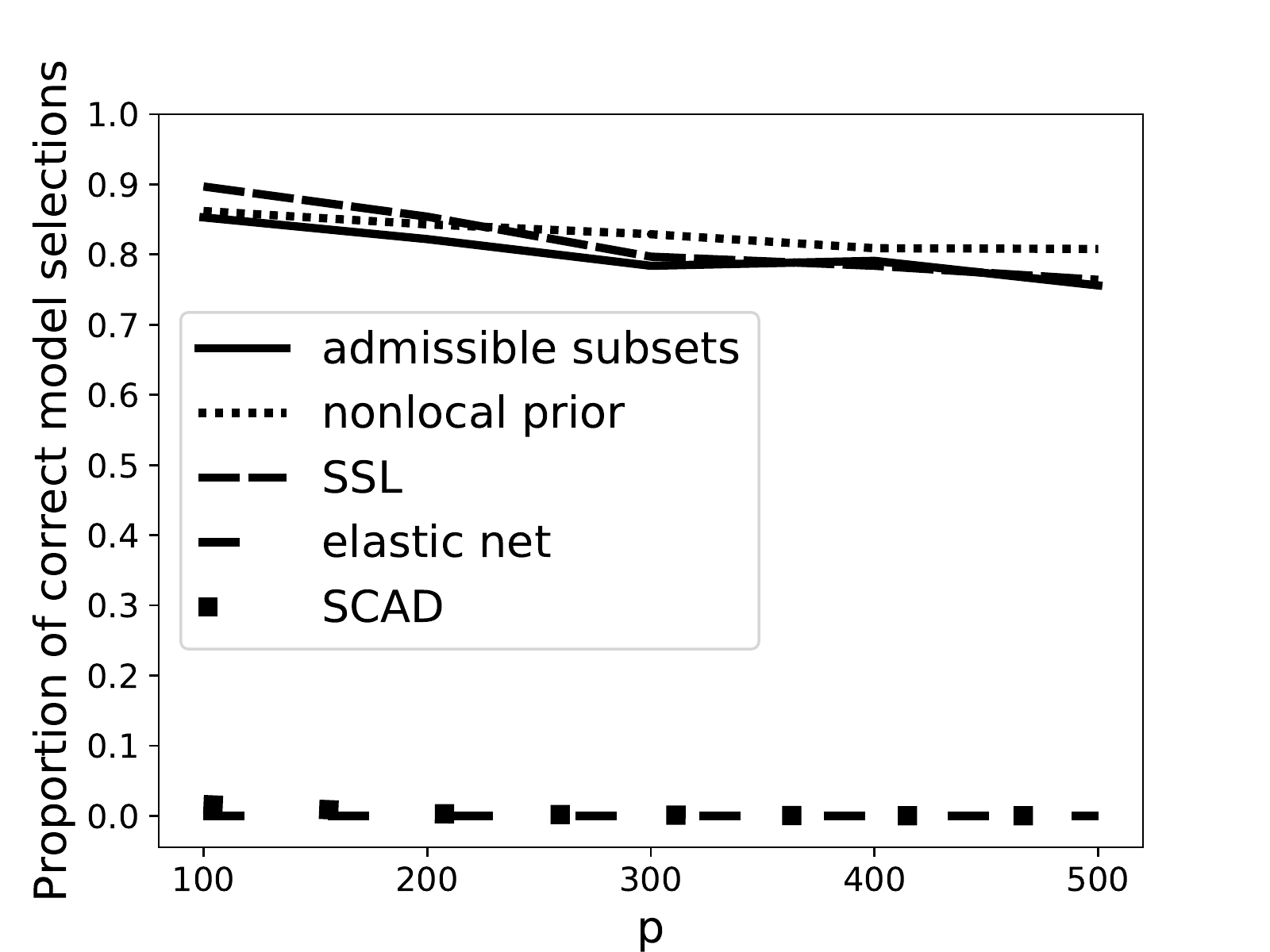}\includegraphics[trim=0cm 0cm 0cm 0cm, clip=true, scale=.38]{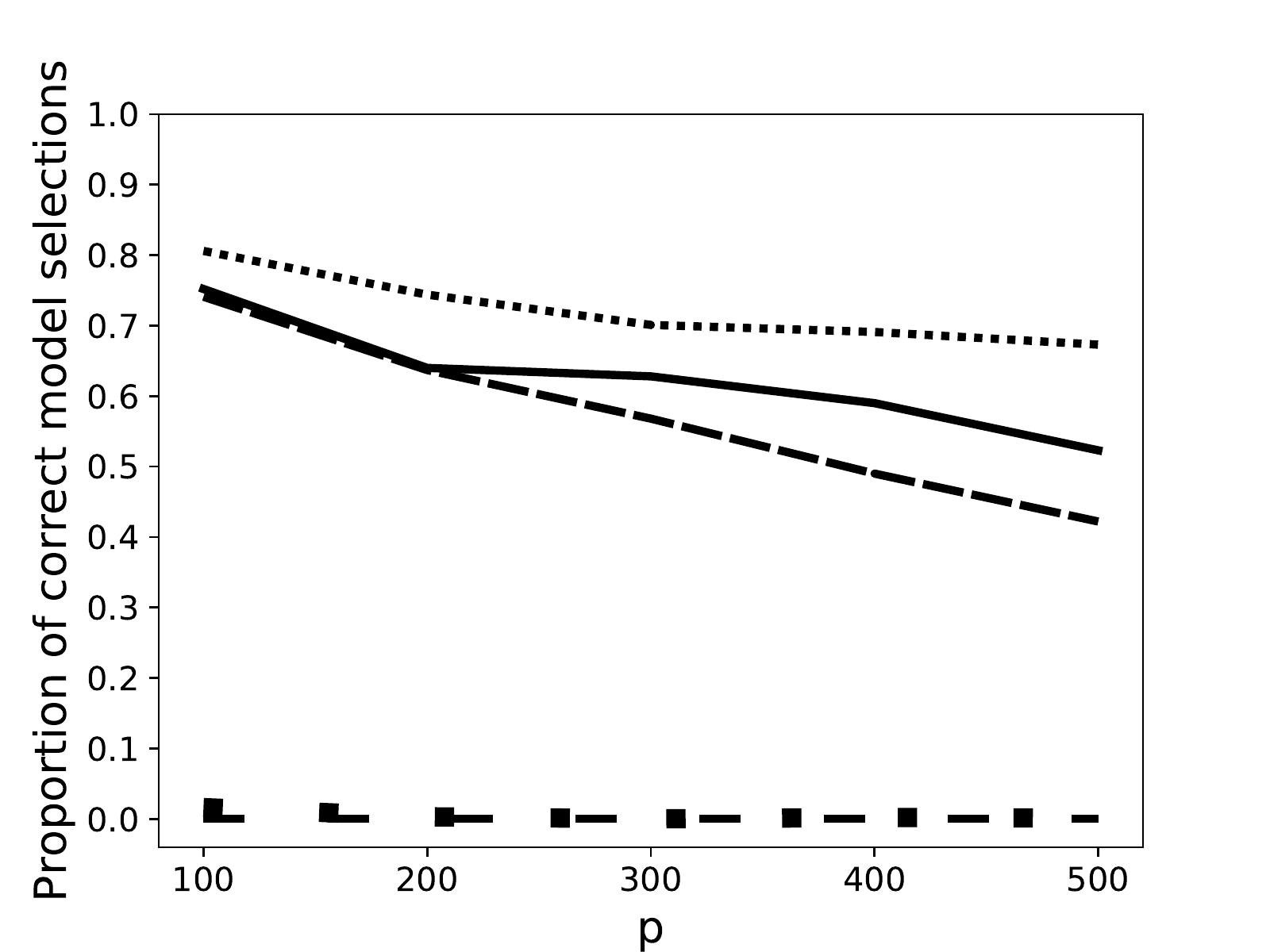}
\includegraphics[trim=0cm 0cm 0cm 0cm, clip=true, scale=.38]{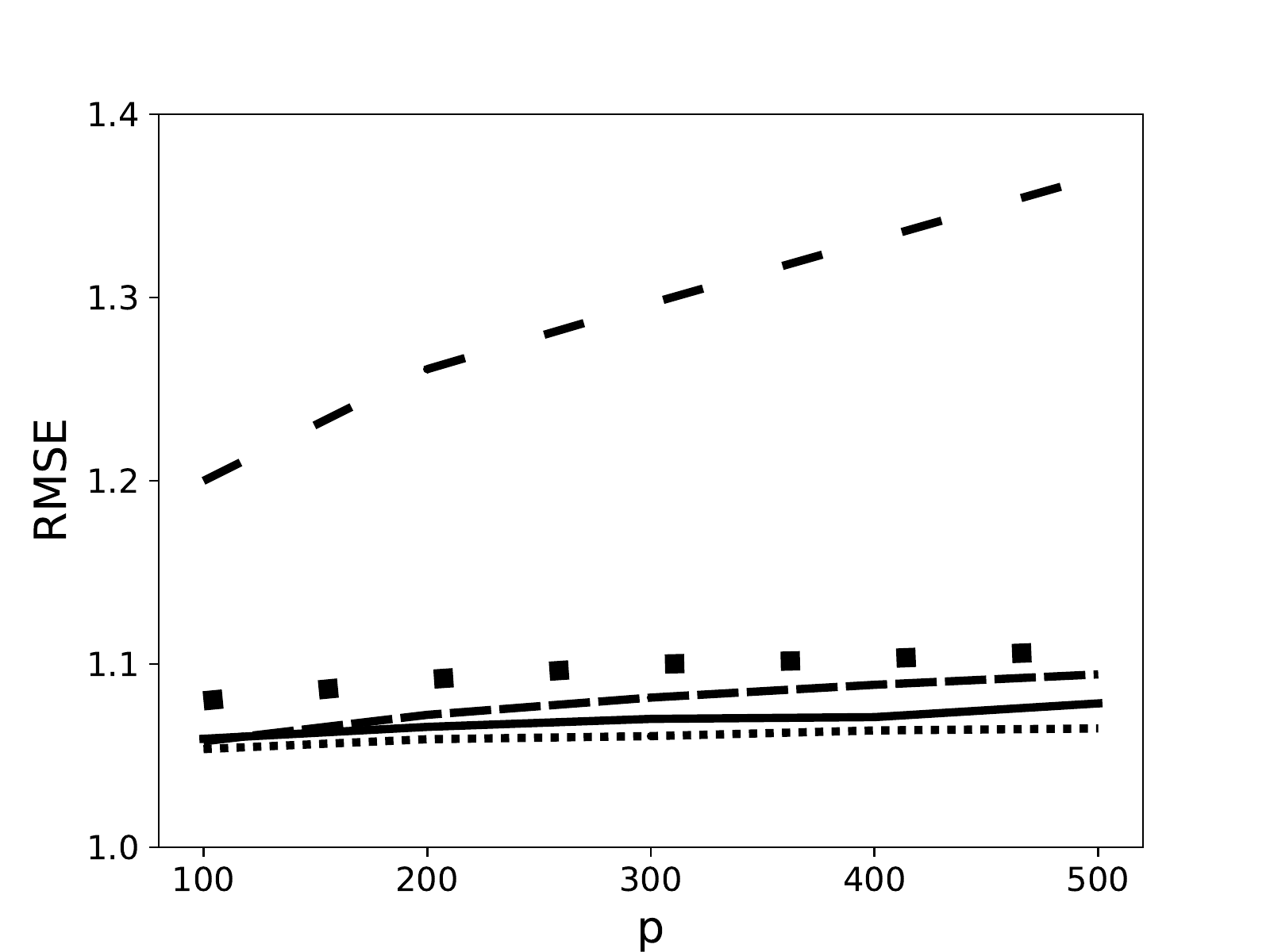}\includegraphics[trim=0cm 0cm 0cm 0cm, clip=true, scale=.38]{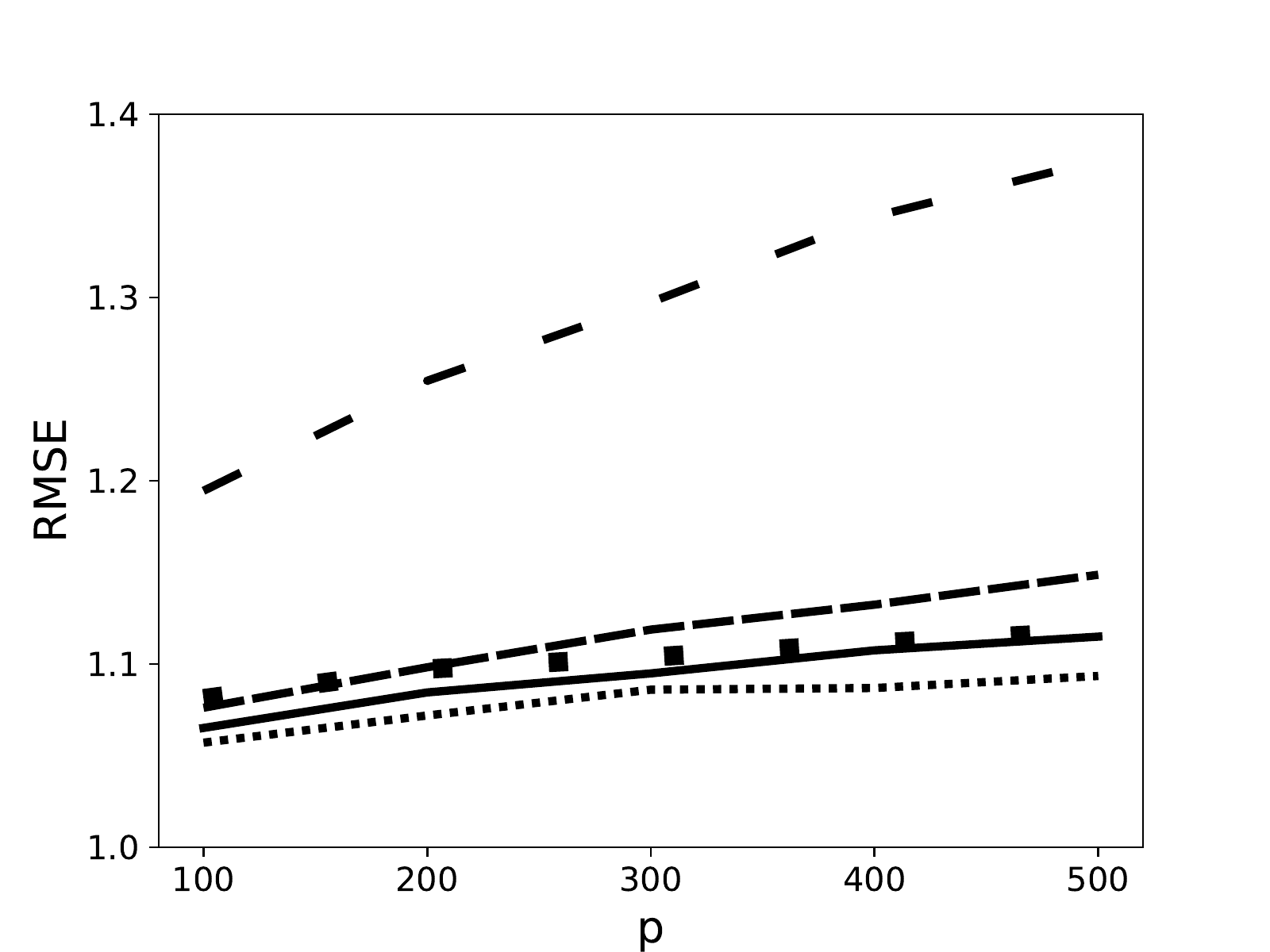}
\vspace{-.1in}
\caption{\footnotesize The average $r(M_{o}|y)$, or average $P(M_{o}|y)$ is displayed in the first row, the average proportion of correct model selections in the second row, and the average RMSE in the third.  Averages are over 1000 synthetic data sets (for $\rho = 0$ and $\rho = .25$, respectively).  For the GFI and Bayesian procedures the MAP subset is used as the estimator of the true model, and in the frequentist procedures the estimated model is considered to be the subset of covariates with nonzero coefficient estimates.  The RMSE is computed on an out-of-sample test set of 100 observations.}\label{output_sim1}
\end{figure}

Figure \ref{output_sim1} shows results of the first simulation setup.  The first row of plots displays the average generalized fiducial probability of the true model (i.e., average $r(M_{o}|y)$), or the average posterior probability of the true model for the Bayesian nonlocal procedure (i.e., average $P(M_{o}|y)$), over the 1000 synthetic data sets (for $\rho = 0$ and $\rho = .25$, respectively).  Conditional on the data, these plots address the consistency of the procedures with respect to the uncertainty from not knowing $M_{o}$.  This generalized fiducial or Bayesian-like consistency is that which is dealt with in Theorem \ref{MainResult}.  Note that frequentist and MAP estimators do not yield posterior probability estimates, and thus cannot be compared to in the first row of plots.

The second row of Figure \ref{output_sim1} shows the average proportion of correct model selections over the 1000 synthetic data sets (for $\rho = 0$ and $\rho = .25$, respectively).  For the GFI and the Bayesian procedures the MAP subset is taken to be the estimator of the true model, and in the frequentist procedures the estimated model is considered to be the subset of covariates with nonzero coefficient estimates.  These plots address the consistency of the procedures with respect to repeated sampling (i.e., frequentist) uncertainty.  Finally, the third row of Figure \ref{output_sim1} presents the average root mean squared error (RMSE) over the 1000 synthetic data sets (for $\rho = 0$ and $\rho = .25$, respectively).  The MAP estimated model is used to compute the RMSE for the GFI and Bayesian procedures.  For all procedures, the RMSE is computed on an out-of-sample test set of 100 observations.

Note that the criterions used in the first two rows of Figure \ref{output_sim1} are very strict.  They only reflect instances when the procedures are exactly correct, and count the procedure as incorrect if it is missing even one covariate from the true model or includes even one spurious covariate.  Often the elastic net and SCAD are able to identify all of the true covariates but estimate extra coefficients to be nonzero.  This results in poor identification of the true subset, and worse out-of-sample prediction error.  The remaining procedures (including our $\varepsilon$-$admissible$ subsets method) struggle to identify the two covariates with smallest coefficient magnitudes, but typically do not introduce more than one or two false positives.  

Our $\varepsilon$-$admissible$ subsets procedure evidently performs the best at assigning highest posterior probability to the true model.  And even in comparison to the frequentist-oriented metrics, proportion of correct model selections and out-of-sample RMSE, our $\varepsilon$-$admissible$ subsets procedure performs more or less on par with the best performing methods considered.  Note that in collinear design settings (i.e., $\rho > 0$) it may be the case that a strict subset of the true data generating model is identified as the `true' model within the $\varepsilon$-$admissible$ framework.  This is meaningful because it manifests the fact that the true model may not be minimal (i.e., contains redundant predictors) in collinear, finite-sample settings.  Furthermore, it explains the larger difference in proportion of true model selections between the $\varepsilon$-$admissible$ and nonlocal prior performances in the $\rho = .25$ case (versus the $\rho = 0$ case), which is accompanied by only a very small change in the difference between RMSE performance.  This phenomenon is illustrated in a more extreme case of collinearity in the next simulation setup.

\subsection{Simulation setup 2}\label{SimResults_setup2}

The $\varepsilon$-$admissible$ subsets approach has been developed in this paper as a method of obtaining a posterior-like distribution which effectively eliminates (i.e., assigns negligible probability to) all subsets with redundancies.  To illustrate that our constructed methods do just that, consider the following setup in which the true data generating model lacks uniqueness for the small sample size $n = 30$:
\vspace{-.18cm}\begin{equation}\small\label{model_setup2}
Y \sim \text{N}_{n}\Big(1\cdot x^{(1)} + 1\cdot x^{(2)} + \cdots + 1\cdot x^{(9)}, I_{n}\Big),
\vspace{-.18cm}\end{equation}
where $x^{(1)}, x^{(2)}, x^{(3)} \overset{\text{iid}}{\sim} \text{N}_{n}(0, I_{n})$, and 
\begin{center}
\small
\begin{tabular}{rrrrrrrrr}
$x^{(4)}$ & $\sim$ & N$_{n}\Big($ & $.25\cdot x^{(1)}$ &        &                            &        &                              & , $.1^{2} I_{n}\Big)$ \\
$x^{(5)}$ & $\sim$ & N$_{n}\Big($ &                              &        & $.5\cdot x^{(2)}$ &        &                              & , $.1^{2} I_{n}\Big)$ \\
$x^{(6)}$ & $\sim$ & N$_{n}\Big($ &                              &        &                            & $-$  & $.75\cdot x^{(3)}$ & , $.1^{2} I_{n}\Big)$ \\
$x^{(7)}$ & $\sim$ & N$_{n}\Big($ &$x^{(1)}$                &        &                            & $+$ & $x^{(3)}$               & , $.1^{2} I_{n}\Big)$ \\
$x^{(8)}$ & $\sim$ & N$_{n}\Big($ &                              &        & $x^{(2)}$             & $-$  & $x^{(3)}$               & , $.1^{2} I_{n}\Big)$ \\
$x^{(9)}$ & $\sim$ & N$_{n}\Big($ &$x^{(1)}$                & $+$ & $x^{(2)}$             & $+$ & $x^{(3)}$               & , $.1^{2} I_{n}\Big)$ \\
\end{tabular}
\end{center}
With standard deviations of .1 and a model error standard deviation of 1, covariates $x^{(4)}, \dots, x^{(9)}$ can all be approximately expressed as a linear combination of $x^{(1)}, x^{(2)}, x^{(3)}$.  Accordingly, with a small increase in error variance, model (\ref{model_setup2}) can be re-expressed using various combinations of the 9 predictors.  However, observe that a subset with 4 or more predictors predominantly contains redundant information.  

Recall from Section \ref{Intro} that the nonlocal prior approach of \cite{Johnson2012} is designed to assign negligible probabilities to subsets containing predictor(s) with coefficients of zero.  So, in theory, the full subset $\{x^{(1)},\dots,x^{(9)}\}$ will remain the best candidate for the true model within the nonlocal prior framework.  In fact, this is demonstrated to be the case in Table \ref{output_sim2} which shows the performance of both the nonlocal prior and the $\varepsilon$-$admissible$ subsets approach on 1000 data vectors $y$ generated according to (\ref{model_setup2}), with each covariate having a `true' coefficient of 1.  Note that as in the first simulation setup 1000 design matrices $X$ are generated to generate the 1000 $y$ vectors.

\begin{table}[H]
\vspace{-.1in}
\centering
\begin{tabular}{r|rrr}
                               & model size & RMSE & $r(M_{\text{MAP}} | y)$ or $P(M_{\text{MAP}} | y)$ \\
\hline 
$\varepsilon$-$admissible$ subsets & 3.476         & 1.138  & .365 \\
nonlocal prior          & 8.997         & 1.197  & .333 \\
\end{tabular}\caption{\footnotesize The average number of covariates in the MAP estimator, $M_{\text{MAP}}$, ($|M_{\text{MAP}}|$) is presented in the first column, the average RMSE in the second, and the average $r(M_{\text{MAP}} | y)$ or $P(M_{\text{MAP}} | y)$ in the third.  Averages are over 1000 synthetic data sets from model (\ref{model_setup2}).  The RMSE is computed on an out-of-sample test set of 30 observations.}\label{output_sim2}  
\vspace{-.2in}
\end{table}

Table \ref{output_sim2} shows that the MAP estimate for the $\varepsilon$-$admissible$ subsets approach contains 3-4 covariates, on average, and that in fact the average RMSE is smaller than that of the nonlocal prior approach.  Indeed, the MAP estimates for the nonlocal prior procedure typically includes all 9 covariates even though the $y$ vectors can be mostly explained by only 3 of the predictors.  This simple simulation illustrates a pivotal difference between the nonlocal prior and $\varepsilon$-$admissible$ subsets approaches.  With $p = 9$ the implication of not discriminating against redundant subsets may seem trivial.  However, the $2^{p}$ size of the sample space grows rapidly in $p$ and thus, puts exponentially more burden on procedures which do not discriminate based on redundancies.  This is reflected by comparing the differences in strength of asymptotic consistency achieved for the two procedures.  Though, the consistency result of the nonlocal prior method from \cite{Johnson2012} is argued to be stronger in an as-of-yet unpublished manuscript by \cite{Shin2015}.

\section{Concluding remarks}

In this paper we have developed a new perspective for variable selection to exploit a non-redundancy property of a true data generating model.  The basic idea calls for defining a true model as one which contains minimal amount of information necessary for explaining and/or predicting the observed data.  The difference between our definition of a true model and the usual definition arises only in finite-sample applications, and was illustrated in Section \ref{SimResults_setup2}.  Within our variable selection framework, this definition allows us to show that the posterior-like probability of the true model converges to 1 asymptotically, even with $p$ growing almost exponentially in $n$, with the intuition that redundancies in the sample space are very effectively eliminated.  Moreover, our empirical simulation results are consistent with this strong consistency result, and as desired, it is demonstrated in a situation of high collinearity that the $\varepsilon$-$admissible$ subsets approach yields a posterior-like distribution which is concentrated over subsets with fewer covariates, without sacrificing prediction error. 

A non-redundancy property of a true data generating model is seemingly general enough to extend to variable or feature selection problems beyond the linear model setting, but would seem infeasible if it could not be developed for the linear model setting.  Thus, the goal of this paper has been to establish the potential feasibility of exploiting such a property.  In future work we hope to extend our methods to more complicated settings.


\bibliographystyle{imsart-nameyear}
\bibliography{References}

\appendix
\section{Proofs}\label{AppendProofs}

{\noindent \bf Proof of Lemma \ref{Term1Bound}.}
From the distributional assumption on $\beta_{M}$ it follows that 
\vspace{-.18cm}\[\small
T := \sqrt{\frac{n-|M|}{\RSS_{M}}}(X_{M}'X_{M})^{\frac{1}{2}}(\beta_{M} - \widehat{\beta}_{M}) \sim \text{t}_{n-|M|}(0,I_{|M|}).
\vspace{-.18cm}\]
Thus, 
\vspace{-.18cm}\[\small
\|X'X_{M}(\beta_{M} - \widehat{\beta}_{M})\|_{2}^{2} = \|AT\|_{2}^{2}\frac{\RSS_{M}}{n-|M|} = \|QDW'T\|_{2}^{2}\frac{\RSS_{M}}{n-|M|}
\vspace{-.18cm}\]
where $A = X'X_{M}(X_{M}'X_{M})^{-\frac{1}{2}}$ is a $p\times |M|$ matrix which has the singular value decomposition $A = QDW'$ for $Q$ and $W$ each orthogonal matrices.  By the definition of the multivariate T distribution, 
\vspace{-.18cm}\[\small
T = \sqrt{\frac{n-|M|}{V}}Z \ \implies \ W'T = \sqrt{\frac{n-|M|}{V}}W'Z =: \sqrt{\frac{n-|M|}{V}}\widetilde{Z} =: \widetilde{T},
\vspace{-.18cm}\]
where $V \sim \chi_{n-|M|}^{2}$ and $Z,\widetilde{Z} \sim \text{MVN}(0,I_{|M|})$.  Then
\vspace{-.18cm}\[\small
\|AT\|_{2}^{2} = \|QD\widetilde{T}\|_{2}^{2} = \widetilde{T}'D'D\widetilde{T} = \sum_{1}^{|M|}\widetilde{T}_{i}^{2}\lambda_{i} \le \Lambda_{M}\max{\widetilde{T}_{j}^{2}}
\vspace{-.18cm}\]
where $\Lambda_{M} = \sum_{1}^{|M|}\lambda_{i}$, and $\lambda_{i}$ is the $i$th eigenvalue of $A'A$.  Observe that $\Lambda_{M}$ also has the more intuitive expression $\Lambda_{M} = \tr(A'A) = \tr\big((H_{M}X)'H_{M}X\big)$ which illustrates that it is intimately related to the presence of collinearity amongst the covariates.

Recall that $\widehat{\sigma}_{M}^{2} := \RSS_{M} / (n - |M|)$.  Then
\vspace{-.18cm}\[\small
\begin{split}
P\Big(\frac{1}{2}\|X'X_{M}(\beta_{M} - \widehat{\beta}_{M})\|_{2}^{2} \ge \frac{\varepsilon}{9}\Big) & = P\Big(\frac{1}{2}\|AT\|_{2}^{2}\frac{\RSS_{M}}{n-|M|} \ge \frac{\varepsilon}{9}\Big) \\
& \le P\Big(\frac{1}{2}\Lambda_{M}\max{\widetilde{T}_{j}^{2}}\frac{\RSS_{M}}{n-|M|} \ge \frac{\varepsilon}{9}\Big) \\
& = P\Big(\frac{n-|M|}{V}\max{Z_{j}^{2}}\widehat{\sigma}_{M}^{2} \ge \frac{2\varepsilon}{9\Lambda_{M}}\Big). \\
\end{split}
\vspace{-.18cm}\]
Since $V \sim \chi_{n-|M|}^{2}$,
\vspace{-.18cm}\begin{equation}\small\label{bound1}
\begin{split}
& P\Big(\frac{1}{2}\|X'X_{M}(\beta_{M} - \widehat{\beta}_{M})\|_{2}^{2} \ge \frac{\varepsilon}{9}\Big) \\
& \ \ \ \ \ \ \ \ \ \ \le \int_{0}^{\infty}P\Big(\max{|Z_{j}|} \ge \frac{\sqrt{2\varepsilon v / (n-|M|)}}{3\widehat{\sigma}_{M}\sqrt{\Lambda_{M}}} \Big) \frac{v^{\frac{n-|M|}{2}-1}e^{-\frac{v}{2}}}{2^{\frac{n-|M|}{2}}\Gamma(\frac{n-|M|}{2})} \ dv \\
& \ \ \ \ \ \ \ \ \ \ \le \int_{0}^{\infty}\sum_{j=1}^{|M|}P\Big(|Z_{j}| \ge \frac{\sqrt{2\varepsilon v / (n-|M|)}}{3\widehat{\sigma}_{M}\sqrt{\Lambda_{M}}} \Big) \frac{v^{\frac{n-|M|}{2}-1}e^{-\frac{v}{2}}}{2^{\frac{n-|M|}{2}}\Gamma(\frac{n-|M|}{2})} \ dv \\
& \ \ \ \ \ \ \ \ \ \ \le \frac{3|M|\widehat{\sigma}_{M}\sqrt{\Lambda_{M}(n - |M|)}}{\sqrt{2\pi\varepsilon}\left(1 + \frac{\varepsilon}{9\Lambda_{M}\widehat{\sigma}_{M}^{2}(\frac{n - |M|}{2})}\right)^{\frac{n-|M|}{2} - \frac{3}{2}}} \cdot\frac{\Gamma(\frac{n-|M|-1}{2})}{\Gamma(\frac{n-|M|}{2})} \\
\end{split}
\vspace{-.18cm}\end{equation}
where the last inequality follows because for the standard normal CDF, $\Phi$ for $x > 0$,
\vspace{-.18cm}\[\small
\Phi(-x) \le \frac{\frac{1}{\sqrt{2\pi}}e^{\frac{-x^{2}}{2}}}{x}.
\vspace{-.18cm}\]

To simplify the bound, observe first that \cite{Jameson2013} gives  
\vspace{-.18cm}\begin{equation}\small\label{obs1}
\frac{\Gamma(\frac{n-|M|-1}{2})}{\Gamma(\frac{n-|M|}{2})} \le \frac{\sqrt{2(n-|M|)}}{n-|M|-1}.
\vspace{-.18cm}\end{equation}
Second, observe that for $0 \le x \le n$,
\vspace{-.18cm}\[\small
\Big(1 + \frac{x}{n}\Big)^{-n} \le e^{-x + \frac{x^{2}}{2n}} = e^{-x(1 - \frac{x}{2n})} \le e^{-\frac{x}{2}}.
\vspace{-.18cm}\]
By assumption $|M| \le c_{1}n$, and $\frac{\varepsilon}{9\Lambda_{M}\widehat{\sigma}_{M}^{2}} < \frac{n-|M|}{2}$ which implies that 
\vspace{-.18cm}\begin{equation}\small\label{obs2}
\left(1 + \frac{\varepsilon}{9\Lambda_{M}\widehat{\sigma}_{M}^{2}(\frac{n - |M|}{2})}\right)^{-\frac{n-|M|}{2} + \frac{3}{2}} \le 2^{\frac{3}{2}}e^{-\frac{\varepsilon}{18\Lambda_{M}\widehat{\sigma}_{M}^{2}}}. 
\vspace{-.18cm}\end{equation}
Therefore, applying (\ref{obs1}) and (\ref{obs2}) to (\ref{bound1}) gives
\vspace{-.18cm}\[\small
P\Big(\frac{1}{2}\|X'X_{M}(\beta_{M} - \widehat{\beta}_{M})\|_{2}^{2} \ge \frac{\varepsilon}{9}\Big) \le \frac{2^{\frac{3}{2}}3|M|\widehat{\sigma}_{M}\sqrt{\Lambda_{M}}}{\sqrt{\pi\varepsilon}(1 - \frac{1}{n-|M|})}e^{-\frac{\varepsilon}{18\Lambda_{M}\widehat{\sigma}_{M}^{2}}}.
\vspace{-.18cm}\]
$\hfill \blacksquare$

{\noindent \bf Proof of Lemma \ref{Term2Bound}.}
Similar to the proof of Lemma \ref{Term1Bound}.
$\hfill \blacksquare$ 

{\noindent \bf Proof of Theorem \ref{BigModel}.}
Recall that 
\vspace{-.18cm}\[\small
h(\beta_{M}) = \I\Big\{\frac{1}{2}\|X'(X_{M}\beta_{M} - Xb_{\min})\|_{2}^{2} \ge \varepsilon \Big\},
\vspace{-.18cm}\]
where $b_{\min}$ solves $\ds \min_{b\in\R^{p}} \frac{1}{2}\|X'(X_{M}\beta_{M} - Xb)\|_{2}^{2}$ subject to $\|b\|_{0} \le |M|-1$.  Observe that
\vspace{-.18cm}\begin{equation}\small\label{EhBoundStart}
\begin{split}
\E(h(\beta_{M})) & = P\Big(\frac{1}{2}\|X'(X_{M}\beta_{M} - Xb_{\min})\|_{2}^{2} \ge \varepsilon\Big) \\
& \le P\Big(\frac{1}{2}\|X'\big(X_{M}\beta_{M} - X\E_{y}(\widehat{\beta}_{M(-1)})\big)\|_{2}^{2} \ge \varepsilon\Big), \\
\end{split}
\vspace{-.18cm}\end{equation}
where $\widehat{\beta}_{M(-1)}$ is the least squares estimate corresponding to the subset of covariates $M$ with one covariate removed so that $\|\E_{y}(\widehat{\beta}_{M(-1)})\|_{0} \le |M| - 1$.  The covariate removed is chosen to correspond to the smallest (in magnitude) component of $\widehat{\beta}_{M}$.  To refine the bound on $\E(h(\beta_{M}))$, decompose the last expression in (\ref{EhBoundStart}) using the triangle inequality as follows.
\vspace{-.18cm}\begin{equation}\small\label{threeTermBound}
\begin{split}
& P\Big(\frac{1}{2}\|X'(X_{M}\beta_{M} - X_{M(-1)}\E_{y}(\widehat{\beta}_{M(-1)}))\|_{2}^{2} \ge \varepsilon\Big) \\
& \ \ \ \ \ \ \ \ \ \ \ \ \ \ \le P\Big(\frac{1}{2}\|X'X_{M}(\beta_{M} - \widehat{\beta}_{M})\|_{2}^{2} \ge \frac{\varepsilon}{9}\Big) \\
& \ \ \ \ \ \ \ \ \ \ \ \ \ \ \ \ \ \ \ \ \ \ \ \ + \I\Big\{\frac{1}{2}\|X'X_{M}\big(\widehat{\beta}_{M} - \E_{y}(\widehat{\beta}_{M})\big)\|_{2}^{2} \ge \frac{\varepsilon}{9}\Big\}, \\
& \ \ \ \ \ \ \ \ \ \ \ \ \ \ \ \ \ \ \ \ \ \ \ \ \ \ \ \ \ \ \  + \I\Big\{\frac{1}{2}\|X'(H_{M} - H_{M(-1)})X_{M_{o}}\beta_{M_{o}}^{0}\|_{2}^{2} \ge \frac{\varepsilon}{9}\Big\}, \\
\end{split}
\vspace{-.18cm}\end{equation}
where $H_{M} := X_{M}(X_{M}'X_{M})^{-1}X_{M}'$.  Observe that by Lemma \ref{Term1Bound} the first term on the right hand side,
\vspace{-.18cm}\[\small
P\Big(\frac{1}{2}\|X'X_{M}(\beta_{M} - \widehat{\beta}_{M})\|_{2}^{2} \ge \frac{\varepsilon}{9}\Big) \le \frac{2^{\frac{3}{2}}3|M|\widehat{\sigma}_{M}\sqrt{\Lambda_{M}}}{\sqrt{\pi\varepsilon}(1 - \frac{1}{n-|M|})}e^{-\frac{\varepsilon}{18\Lambda_{M}\widehat{\sigma}_{M}^{2}}}.
\vspace{-.18cm}\]
Note that the last two terms are written as an indicator function because the uncertainty here comes from $\beta_{M}$.  However, the second term results from the uncertainty in observing $Y$.  Accordingly, by Lemma \ref{Term2Bound},
\vspace{-.18cm}\[\small
P_{y}\Big(\I\Big\{\frac{1}{2}\|X'X_{M}\big(\widehat{\beta}_{M} - \E_{y}(\widehat{\beta}_{M})\big)\|_{2}^{2} \ge \frac{\varepsilon}{9}\Big\} = 0\Big) \ge 1 - \frac{3|M|\sigma_{M}\sqrt{\Lambda_{M}}}{\sqrt{\varepsilon\pi}e^{\frac{\varepsilon}{9\sigma^{2}_{M}\Lambda_{M}}}}. 
\vspace{-.18cm}\]
Recall that $P_{y}$ is used to denote the probability measure associated with the sampling distribution of $Y$.  The third term in (\ref{threeTermBound}) must be zero by Condition \ref{condition3}.
$\hfill \blacksquare$

{\noindent \bf Proof of Theorem \ref{TrueModelCase}.}
Recall that 
\vspace{-.18cm}\[\small
h(\beta_{M_{o}}) = \I\Big\{\frac{1}{2}\|X'(X_{M_{o}}\beta_{M_{o}} - Xb'_{\min})\|_{2}^{2} \ge \varepsilon \Big\},
\vspace{-.18cm}\]
where $b'_{\min}$ solves $\ds \min_{b\in\R^{p}} \frac{1}{2}\|X'(X_{M_{o}}\beta_{M_{o}} - Xb)\|_{2}^{2}$ subject to $\|b\|_{0} \le |M_{o}|-1$.  To show the desired result, let $b_{\min}$ be the solution to $\ds \min_{b\in\R^{p}} \|X'(X_{M_{o}}\beta_{M_{o}}^{0} - Xb)\|_{2}^{2}$ subject to $\|b\|_{0} \le |M_{o}|-1$.  Then observe that
\vspace{-.18cm}\[\small
\begin{split}
\|X'(X_{M_{o}}\beta_{M_{o}}^{0} - Xb_{\min})\|_{2} & \le \|X'(X_{M_{o}}\beta_{M_{o}}^{0} - Xb'_{\min})\|_{2} \\
& \le \|X'X_{M_{o}}(\beta_{M_{o}}^{0} - \beta_{M_{o}})\|_{2} \\
& \ \ \ \ \ \ \ \ \ \ \ \ \ \ \ + \|X'(X_{M_{o}}\beta_{M_{o}} - Xb'_{\min})\|_{2}. \\
\end{split}
\vspace{-.18cm}\]
Note the difference between $b_{\min}$ and $b'_{\min}$ here.  The rightmost term is the quantity of interest because it will become $\E(h(\beta_{M_{o}}))$ in the next few steps.  The term on the left of the inequality corresponds to the quantity in Condition \ref{condition1}.

Adding and subtracting $\widehat{\beta}_{M_{o}}$ inside the first term on the right side of the second inequality, and applying the triangle inequality gives,
\vspace{-.18cm}\[\small
\begin{split}
& \I\Big\{\frac{1}{2}\|X'(X_{M_{o}}\beta_{M_{o}}^{0} - Xb_{\min})\|_{2}^{2} \ge 9\varepsilon\Big\} \\
& \ \ \ \ \ \ \ \ \ \ \ \ \ \ \ \le P\Big(\frac{1}{2}\|X'X_{M_{o}}(\beta_{M_{o}} - \widehat{\beta}_{M_{o}}\|_{2}^{2} \ge \varepsilon\Big) \\
& \ \ \ \ \ \ \ \ \ \ \ \ \ \ \ \ \ \ \ \ \ + \I\Big\{\frac{1}{2}\|X'X_{M_{o}}(\widehat{\beta}_{M_{o}} - \E_{y}(\widehat{\beta}_{M_{o}})\|_{2}^{2} \ge \varepsilon\Big\} + \E(h(\beta_{M_{o}})), \\
\end{split}
\vspace{-.18cm}\]
and by applying Lemma \ref{Term1Bound},
\vspace{-.18cm}\[\small
\begin{split}
& \I\Big\{\frac{1}{2}\|X'(X_{M_{o}}\beta_{M_{o}}^{0} - Xb_{\min})\|_{2}^{2} \ge 9\varepsilon\Big\} \\
& \ \ \ \ \ \ \ \ \ \ \ \ \ \ \ \le \frac{2^{\frac{3}{2}}3p_{M_{o}}\widehat{\sigma}_{M_{o}}\sqrt{\Lambda_{M_{o}}}}{\sqrt{\pi\varepsilon}(1 - \frac{1}{n-p_{M_{o}}})}e^{-\frac{\varepsilon}{18\Lambda_{M_{o}}\widehat{\sigma}_{M_{o}}^{2}}} \\
& \ \ \ \ \ \ \ \ \ \ \ \ \ \ \ \ \ \ \ \ \ + \I\Big\{\frac{1}{2}\|X'X_{M_{o}}(\widehat{\beta}_{M_{o}} - \E_{y}(\widehat{\beta}_{M_{o}})\|_{2}^{2} \ge \frac{\varepsilon}{9}\Big\} + \E(h(\beta_{M_{o}})) \\
\end{split}
\vspace{-.18cm}\]
where the middle term is written as an indicator function because the uncertainty here comes from $\beta_{M_{o}}$.  This indicator is 0 by Lemma \ref{Term2Bound} with probability exceeding (\ref{aaa}).  Therefore, since Condition \ref{condition1} implies that the indicator on the left side of the inequality is 1,
\vspace{-.18cm}\[\small
1 - \frac{2^{\frac{3}{2}}3p_{M_{o}}\widehat{\sigma}_{M_{o}}\sqrt{\Lambda_{M_{o}}}}{\sqrt{\pi\varepsilon}(1 - \frac{1}{n-p_{M_{o}}})}e^{-\frac{\varepsilon}{18\Lambda_{M_{o}}\widehat{\sigma}_{M_{o}}^{2}}} \le \E(h(\beta_{M_{o}}))
\vspace{-.18cm}\]
with probability exceeding 
\vspace{-.18cm}\begin{equation}\small\label{aaa}
1 - \frac{3p_{M_{o}}\sigma_{M_{o}}\sqrt{\Lambda_{M_{o}}}}{\sqrt{\varepsilon\pi}}e^{-\frac{\varepsilon}{9\sigma_{M_{o}}^{2}\Lambda_{M_{o}}}},
\vspace{-.18cm}\end{equation}
due to the uncertainty in observing $Y$.
$\hfill \blacksquare$

The following lemma is needed for the proof of the main result, Theorem \ref{MainResult}. 

\begin{lemma}\label{RSSLemma}
Assume all conditions and notations of Theorem \ref{MainResult}, and without loss of generality suppose $\sigma_{M_{o}}^{0} = 1$, where $\sigma_{M_{o}}^{0}$ is the true but unknown error standard deviation.  Then the following holds.
\begin{description}
	\item[Case 1] This case pertains to subsets $M$ with $|M| \le |M_{o}|$.
		\vspace{-.18cm}\[\small
		P_{y}\left(\bigcap_{j=1}^{|M_{o}|}\bigcap_{\mathcal{M}_{j}}\Big\{Y : \Big(\frac{\RSS_{M_{o}}}{\RSS_{M}}\Big)^{\frac{n-j-1}{2}} \le e^{-2|M_{o}|\log(p)}\Big\}\right) \ge 1 - V_{1},
		\vspace{-.18cm}\]
		where $\mathcal{M}_{j} := \{M \ne M_{o}: |M| = j\}$,
		\vspace{-.18cm}\[\small
		\begin{split}
		V_{1} := & \max_{\substack{M\ne M_{o} \\ |M|\le |M_{o}|}}\Bigg\{ \frac{12|M_{o}|e^{\frac{-\Delta_{M}}{72} + |M_{o}|\log(p)}}{\sqrt{2\pi\Delta_{M}}} + |M_{o}|e^{-\frac{\Delta_{M}}{12} + \frac{|M_{o}|}{2} + |M_{o}|\log(p)} \\
		& \ \ \ \ \ \ \ \ \ \ \ \ \ \ \ \ \ \ \ + |M_{o}|e^{-\frac{\xi_{n,|M_{o}|}}{48}\frac{(n-|M_{o}|-1)\Delta_{M}}{\log(n)^{\gamma}\log(p)} + \frac{n-|M_{o}|}{2} + |M_{o}|\log(p)}\Bigg\}, \\
		\end{split}
		\vspace{-.18cm}\]
		and $\xi_{n,j} = 1 - \frac{2\log(n)^{\gamma}\log(p)}{(n-j-1)/2} \to 1$.

	\item[Case 2] This case pertains to subsets $M$ with $|M_{o}|+1 \le |M| \le n^{\alpha}$.
		\vspace{-.18cm}\[\small
		P_{y}\left(\bigcap_{j=|M_{o}|+1}^{n^{\alpha}}\bigcap_{M:|M|=j}\Big\{Y : \Big(\frac{\RSS_{M_{o}}}{\RSS_{M}}\Big)^{\frac{n-j-1}{2}} \le e^{e^{2}(n^{\alpha} + j\log(p))}\Big\}\right) \ge 1 - V_{2},
		\vspace{-.18cm}\]
		where
		\vspace{-.18cm}\[\small
		\begin{split}
		V_{2} & := n^{\alpha} e^{-b_{n}\big[2n^{\alpha} - \frac{|M_{o}|}{2b_{n}} + \big(2 - \frac{1}{b_{n}} - \frac{1}{2b_{n}\log(p)} \big)(|M_{o}|+1)\log(p) \big]} \\
    		& \ \ \ \ \ \ \ \ \ \ \ \ \ \ \ \ \ \ \ \ \ \ \ \ \ \ \ \ \ \ \ \ \ \ \ \ \ \ + \frac{ n^{\alpha}e^{-\frac{n-n^{\alpha}-|M_{o}|}{2} + n^{\alpha}\log(p)} }{ \sqrt{\pi(n-n^{\alpha}-|M_{o}|)} },  \\
    		\end{split}
    	        \vspace{-.18cm}\]
    		and $b_{n} := \frac{n-n^{\alpha}-|M_{o}|}{n-|M_{o}|-2} \to 1$.
\end{description}
\end{lemma}
Note that $V_{1}$ and $V_{2}$ both vanish for large $n$ and $p$ by Condition \ref{condition2}.  This condition also ensures that $\xi_{n,j} \in (0,1)$ which is needed in the proof of this lemma. 

{\noindent \bf Proof of Lemma \ref{RSSLemma}.}
There are two cases to consider.
\begin{description}
 \item[Case 1]  For $M \in \mathcal{M}_{j}$ with $j \le |M_{o}|$ let $\xi_{n,j} = 1 - \frac{2\log(n)^{\gamma}\log(p)}{(n-j-1)/2} \in (0,1)$ by Condition \ref{condition2} for large $n$ and $p$.  Then 
    \vspace{-.18cm}\begin{equation}\small\label{InitialRelations}
    \begin{split}
    & P_{y}\left(\Big(\frac{\RSS_{M_{o}}}{\RSS_{M}}\Big)^{\frac{n-j-1}{2}} > \xi_{n,j}^{\frac{n-j-1}{2}}\right) \\
    & = P_{y}\big(U'(I_{n} - H_{M_{o}})U/\xi_{n,j} > \Delta_{M} + 2\sqrt{\Delta_{M}}Z + U'(I_{n} - H_{M})U\big) \\
    \end{split}
    \vspace{-.18cm}\end{equation}
    since by assumption $Y = X_{M_{o}}\beta_{M_{o}}^{0} + U$ with $U \sim \text{N}_{n}(0,I_{n})$, and so
    \vspace{-.18cm}\[\small
    \begin{split}
    \RSS_{M_{o}} & = U'(I_{n} - H_{M_{o}})U \ \ \text{and} \\
    \RSS_{M} & = \Delta_{M} + 2\beta_{M_{o}}^{0'}X'_{M_{o}}(I_{n} - H_{M})U + U'(I_{n} - H_{M})U \\
    & = Z_{\Delta_{M}} + U'(I_{n} - H_{M})U \\
    \end{split}
    \vspace{-.18cm}\]
    where $Z_{\Delta_{M}} := \Delta_{M} + 2\sqrt{\Delta_{M}}Z$, and $Z \sim \text{N}(0,1)$.  Recall that $\Delta_{M} := \beta_{M_{o}}^{0'}X'_{M_{o}}(I_{n} - H_{M})X_{M_{o}}\beta_{M_{o}}^{0}$.  Continuing in (\ref{InitialRelations}) by subtracting $\chi_{n-|M_{o}|}^{2}$ from both sides of the inequality gives,
    \vspace{-.18cm}\begin{equation}\small\label{case1beginningArgument}
    \begin{split}
    & P_{y}\Big(\frac{\RSS_{M_{o}}}{\RSS_{M}} > \xi_{n,j}\Big) \\
    & \ \ \ \ \ \ \ \ \ \ \ = P_{y}\big( \chi_{n-|M_{o}|}^{2}/\xi_{n,j} - \chi_{n-|M_{o}|}^{2} > Z_{\Delta_{M}} + \chi_{n-j}^{2} - \chi_{n-|M_{o}|}^{2} \big) \\
    & \ \ \ \ \ \ \ \ \ \ \ = P_{y}\big( \chi_{n-|M_{o}|}^{2}(1/\xi_{n,j} - 1) > Z_{\Delta_{M}} + \chi_{|M_{o}|}^{2} - \chi_{j}^{2}\big) \\
    & \ \ \ \ \ \ \ \ \ \ \ \le P_{y}\big( \chi_{n-|M_{o}|}^{2}(1/\xi_{n,j} - 1) > Z_{\Delta_{M}} - \chi_{j}^{2}\big) \\
    \end{split}
    \vspace{-.18cm}\end{equation}
    The last inequality follows because the $\chi_{|M_{o}|}^{2}$ random variable is nonnegative, and removing it simplifies the remaining argument.  Then 
    \vspace{-.18cm}\[\small
    \begin{split}
    & P_{y}\Big(\frac{\RSS_{M_{o}}}{\RSS_{M}} > \xi_{n,j}\Big) \\
    & \le P_{y}\big( \xi_{n,j}\chi_{j}^{2} + \chi_{n-|M_{o}|}^{2}(1 - \xi_{n,j}) - 2\xi_{n,j}\sqrt{\Delta_{M}}Z > \xi_{n,j}\Delta_{M} \big) \\
    &  \le P_{y}\big( |Z| > \sqrt{\Delta_{M}}/6 \big) + P_{y}\big( \chi_{j}^{2} > \Delta_{M}/3 \big) + P_{y}\Big(\chi_{n-|M_{o}|}^{2} > \frac{\xi_{n,j}\Delta_{M}}{3(1 - \xi_{n,j})}\Big). \\
    \end{split}
    \vspace{-.18cm}\]
    For the second and third term, apply the Chernoff bound, and evaluate the moment generating function for the $\chi^{2}_{j}$ and $\chi^{2}_{n-|M_{o}|}$ distributions at $1/4$.  Accordingly,
    \vspace{-.18cm}\[\small
    \begin{split}
    & P_{y}\Big(\frac{\RSS_{M_{o}}}{\RSS_{M}} > \xi_{n,j}\Big) \\
    &  \ \ \ \ \ \ \ \ \ \ \ \le 2 P_{y}\big( Z < -\sqrt{\Delta_{M}}/6 \big) + e^{-\frac{\Delta_{M}}{12} + \frac{j}{2}} + e^{-\frac{\xi_{n,j}\Delta_{M}}{12(1 - \xi_{n,j})} + \frac{n-|M_{o}|}{2}}. \\
    \end{split}
    \vspace{-.18cm}\]
    
    Finally the remaining probability can be controlled by the bound for the CDF of a standard normal random variable for $x > 0$,
    \vspace{-.18cm}\[\small
    \Phi(-x) \le \frac{\frac{1}{\sqrt{2\pi}}e^{\frac{-x^{2}}{2}}}{x}.
    \vspace{-.18cm}\]
    Hence,
    \vspace{-.18cm}\[\small
    \begin{split}
    & P_{y}\Big(\frac{\RSS_{M_{o}}}{\RSS_{M}} > \xi_{n,j}\Big) \\
    & \ \ \ \ \ \ \ \ \ \ \ \le \frac{12e^{\frac{-\Delta_{M}}{72}}}{\sqrt{2\pi\Delta_{M}}} + e^{-\frac{\Delta_{M}}{12} + \frac{|M_{o}|}{2}} + e^{-\frac{\xi_{n,|M_{o}|}}{48}\frac{(n-|M_{o}|-1)\Delta_{M}}{\log(n)^{\gamma}\log(p)} + \frac{n-|M_{o}|}{2}}, \\
    \end{split}
    \vspace{-.18cm}\]
    where the last inequality follows by observing that $j \le |M_{o}|$, and recalling the expression for $\xi_{n,j}$.
    
    Therefore, the probability that $\Big(\frac{\RSS_{M_{o}}}{\RSS_{M}}\Big)^{\frac{n-j-1}{2}} > \xi_{n,j}^{\frac{n-j-1}{2}}$ is satisfied for some $M$ with $|M| \le |M_{o}|$ is
    \vspace{-.18cm}\[\small
    \begin{split}
    & P_{y}\left(\bigcup_{j=1}^{|M_{o}|}\bigcup_{\mathcal{M}_{j}}\left\{\Big(\frac{\RSS_{M_{o}}}{\RSS_{M}}\Big)^{\frac{n-j-1}{2}} > \xi_{n,j}^{\frac{n-j-1}{2}}\right\}\right) \\
    & \ \ \ \ \ \ \ \ \ \ \ \ \ \ \ \ \ \ \ \ \ \le \sum_{j=1}^{|M_{o}|} \binom{p}{j} \max_{\mathcal{M}_{j}}P_{y}\left(\Big(\frac{\RSS_{M_{o}}}{\RSS_{M}}\Big)^{\frac{n-j-1}{2}} > \xi_{n,j}^{\frac{n-j-1}{2}}\right). \\
    \end{split}
    \vspace{-.18cm}\]
    Note that 
    \vspace{-.18cm}\begin{equation}\small\label{binomCoefBehavior}
    \binom{p}{j} = \frac{p(p-1)\cdots(p-j+1)}{j!} = \frac{p^{j}(1-\frac{1}{p})\cdots(1-\frac{j-1}{p})}{j!} \le p^{j}.
    \vspace{-.18cm}\end{equation}
    In fact, \cite{Luo2013} show that if $\log(j)/\log(p) \to \delta$ as $p \to \infty$, for some $\delta > 0$, then $\log\binom{p}{j} = j\log(p)(1-\delta)(1+o(1))$.  Thus,
    \vspace{-.18cm}\[\small
    \begin{split}
    & P_{y}\left(\bigcup_{j=1}^{|M_{o}|}\bigcup_{\mathcal{M}_{j}}\left\{\Big(\frac{\RSS_{M_{o}}}{\RSS_{M}}\Big)^{\frac{n-j-1}{2}} > \xi_{n,j}^{\frac{n-j-1}{2}}\right\}\right) \\
    & \ \ \ \ \ \ \ \ \ \le \sum_{j=1}^{|M_{o}|}  \max_{\mathcal{M}_{j}}\Bigg\{ \frac{12e^{\frac{-\Delta_{M}}{72} + j\log(p)}}{\sqrt{2\pi\Delta_{M}}} + e^{-\frac{\Delta_{M}}{12} + \frac{|M_{o}|}{2} + j\log(p)} \\
    & \ \ \ \ \ \ \ \ \ \ \ \ \ \ \ \ \ \ \ \ \ \ \ \ \ \ \ \ \ \ \ \ + e^{-\frac{\xi_{n,|M_{o}|}}{48}\frac{(n-|M_{o}|-1)\Delta_{M}}{\log(n)^{\gamma}\log(p)} + \frac{n-|M_{o}|}{2} + j\log(p)} \Bigg\}. \\
    \end{split}
    \vspace{-.18cm}\]
    Since $\xi_{n,|M_{o}|} \to 1$, this bound vanishes by Condition \ref{condition2}.  Therefore, with probability exceeding one minus the above bound,
    \vspace{-.18cm}\[\small
    \begin{split}
    \Big(\frac{\RSS_{M_{o}}}{\RSS_{M}}\Big)^{\frac{n-j-1}{2}} & \le \Big(1 - \frac{2\log(n)^{\gamma}\log(p)}{(n-j-1)/2}\Big)^{\frac{n-j-1}{2}} \\
    & \le e^{-2\log(n)^{\gamma}\log(p)} \\
    \end{split}
    \vspace{-.18cm}\]
    uniformly over all $M$ such that $|M| \le |M_{o}|$.

  \item[Case 2]  Consider any subset $M$ with $|M_{o}| < |M| \le n^{\alpha}$ for some positive constant $\alpha < 1$, and let $\{a_{n}\}$ be an arbitrarily sequence of numbers.  To begin, repeating the steps in (\ref{case1beginningArgument}), but subtracting $\chi_{n-j}^{2}/\xi_{n,j}$ on both sides instead of $\chi_{n-|M_{o}|}^{2}$, and replacing the label $\xi_{n,j}$ with $a_{n}$, yields
    \vspace{-.18cm}\[\small
    P_{y}\left(\Big(\frac{\RSS_{M_{o}}}{\RSS_{M}}\Big)^{\frac{n-j-1}{2}} > a_{n}^{\frac{n-j-1}{2}}\right) \le P_{y}\big( \chi_{j}^{2} > a_{n}Z_{\Delta_{M}} + \chi_{n-j}^{2}(a_{n} - 1) \big),
    \vspace{-.18cm}\]
    where $Z_{\Delta_{M}} = \Delta_{M} + 2\sqrt{\Delta_{M}}Z$, $Z \sim N(0,1)$, and $\Delta_{M} = \beta_{M_{o}}^{0'}X'_{M_{o}}(I_{n} - H_{M})X_{M_{o}}\beta_{M_{o}}^{0}$.  Since $M_{o} \subset M$ implies $\Delta_{M} = 0$, the above bound can be simplified by including in the subset $M$ any covariates in $M_{o}$ not already included in $M$.  Accordingly, let $M' := M\cup M_{o}$ which includes $j + l$ covariates, where $l \in\{0,\dots,|M_{o}|\}$ is the number of covariates not shared by $M$ and $M_{o}$.  Then because $\RSS_{M'} \le \RSS_{M}$, 
    \vspace{-.18cm}\[\small
    \begin{split}
    P_{y}\left(\Big(\frac{\RSS_{M_{o}}}{\RSS_{M}}\Big)^{\frac{n-j-1}{2}} > a_{n}^{\frac{n-j-1}{2}}\right) & \le P_{y}\left(\Big(\frac{\RSS_{M_{o}}}{\RSS_{M'}}\Big)^{\frac{n-j-1}{2}} > a_{n}^{\frac{n-j-1}{2}}\right) \\
    & \le P_{y}\big( \chi_{j+l}^{2} > \chi_{n-j-l}^{2}(a_{n} - 1) \big), \\
    \end{split}
    \vspace{-.18cm}\]
    and for any nonnegative $s \in \R$,
    \vspace{-.18cm}\begin{equation}\small\label{bigModelFinalBound}
    \begin{split}
    & P_{y}\Big(\frac{\RSS_{M_{o}}}{\RSS_{M}} > a_{n}\Big) \\
    & \le P_{y}\big(\big\{ \chi_{j+l}^{2} > s(a_{n} - 1) \big\}\cap\big\{ \chi_{n-j-l}^{2} \ge s \big\}\big) \\
    & \ \ \ \ \ \ \ \ \ \ \ \ \ \ \ \ \ \ \ + P_{y}\big(\big\{ \chi_{j+l}^{2} > \chi_{n-j-l}^{2}(a_{n} - 1) \big\}\cap\big\{ \chi_{n-j-l}^{2} < s \big\}\big) \\
    & \le P_{y}\big( \chi_{j+l}^{2} > s(a_{n} - 1) \big) + P_{y}(\chi_{n-j-l}^{2} < s) \\
    \end{split}
    \vspace{-.18cm}\end{equation}
        
    Consider each of these last two terms in turn.  For the first term apply the Chernoff bound, and evaluate the moment generating function for the $\chi^{2}_{j+l}$ distribution at $1/4$.  That gives
    \vspace{-.18cm}\[\small
    P_{y}\big( \chi_{j+l}^{2} > s(a_{n} - 1) \big) \le 2^{\frac{j+l}{2}} e^{-\frac{s(a_{n} - 1)}{4}} \le e^{-\frac{s(a_{n} - 1)}{4} + \frac{j+l}{2}}.
    \vspace{-.18cm}\]
    For the second term in (\ref{bigModelFinalBound}) write out the expression to evaluate the probability explicitly, and then apply the simple bound $e^{-x} \le 1$ for all $x \ge 0$.  Noting that $s > 0$,
    \vspace{-.18cm}\[\small
    P_{y}(\chi_{n-j-l}^{2} < s) \le \frac{1}{ 2^{\frac{n-j-l}{2}} \Gamma\Big(\frac{n-j-l}{2}\Big) } \frac{s^{\frac{n-j-l}{2}}}{\frac{n-j-l}{2}} \le \frac{ (e\cdot s)^{\frac{n-j-l}{2}} \cdot 2^{-\frac{n-j-l}{2}} }{ \sqrt{2\pi} (\frac{n-j-l}{2})^{\frac{n-j-l}{2} + \frac{1}{2}} },
    \vspace{-.18cm}\]
    where the last inequality follows from the well known Sterling lower bound on the gamma function
    \vspace{-.18cm}\[\small
    \Gamma(x) \ge \sqrt{2\pi} x^{x - \frac{1}{2}} e^{-x} \ \ \text{for} \ \ x > 0.
    \vspace{-.18cm}\]
    It is clear from the last expression that for the probability to vanish, $e \cdot s$ must grow no faster than $\frac{n-j-l}{2}$.  Accordingly, choosing $s = \frac{n-j-l}{e^{2}}$ gives
    \vspace{-.18cm}\[\small
    P_{y}\Big(\chi_{n-j-l}^{2} < \frac{n-j-l}{e^{2}}\Big) \le \frac{ e^{-\frac{n-j-l}{2}} }{ \sqrt{\pi(n-j-l)} }. 
    \vspace{-.18cm}\]
    
    Combining the two bounds for (\ref{bigModelFinalBound}) now yields
    \vspace{-.18cm}\[\small
    P_{y}\Big(\frac{\RSS_{M_{o}}}{\RSS_{M}} > a_{n}\Big) \le e^{- (\frac{n-j-l}{e^{2}}) \frac{a_{n} - 1}{4} + \frac{j+l}{2}} + \frac{ e^{-\frac{n-j-l}{2}} }{ \sqrt{\pi(n-j-l)} }.
    \vspace{-.18cm}\]
    It only remains to choose the smallest $a_{n}$ such that the first term in the bound vanishes exponentially fast so that the cumulative probability will vanish in probability over all subsets $M$ with $|M| \le n^{\alpha}$.  Accordingly, it should become apparent shortly that a good choice is 
    \vspace{-.18cm}\begin{equation}\small\label{a_nChoice}
    a_{n} = 1 + \frac{8e^{2}(n^{\alpha} + j\log(p))}{n - j - 1}.
    \vspace{-.18cm}\end{equation}
        
    The probability that $\Big(\frac{\RSS_{M_{o}}}{\RSS_{M}}\Big)^{\frac{n-j-1}{2}} > a_{n}^{\frac{n-j-1}{2}}$ is satisfied for some $M$ with $|M_{o}| < |M| \le n^{\alpha}$ is
    \vspace{-.18cm}\[\small
    \begin{split}
    & P_{y}\left(\bigcup_{j=|M_{o}|+1}^{n^{\alpha}}\bigcup_{M:|M|=j}\left\{\Big(\frac{\RSS_{M_{o}}}{\RSS_{M}}\Big)^{\frac{n-j-1}{2}} > a_{n}^{\frac{n-j-1}{2}}\right\}\right) \\
    & \ \ \ \ \ \ \ \ \ \ \ \ \ \ \ \ \ \le \sum_{j=|M_{o}|+1}^{n^{\alpha}} \binom{p}{j} \max_{M:|M|=j}P_{y}\left(\Big(\frac{\RSS_{M_{o}}}{\RSS_{M}}\Big)^{\frac{n-j-1}{2}} > a_{n}^{\frac{n-j-1}{2}}\right). \\
    \end{split}
    \vspace{-.18cm}\]
    Thus, bounding the binomial coefficient as in (\ref{binomCoefBehavior}), and substituting (\ref{a_nChoice}) for $a_{n}$ yields 
    \vspace{-.18cm}\[\small
    \begin{split}
    & P_{y}\left(\bigcup_{j=|M_{o}|+1}^{n^{\alpha}}\bigcup_{M:|M|=j}\left\{\Big(\frac{\RSS_{M_{o}}}{\RSS_{M}}\Big)^{\frac{n-j-1}{2}} > a_{n}^{\frac{n-j-1}{2}}\right\}\right) \\
    & \le \sum_{j=|M_{o}|+1}^{n^{\alpha}} e^{-b_{n}\big[2n^{\alpha} - \frac{l}{2b_{n}} + \big(2 - \frac{1}{b_{n}} - \frac{1}{2b_{n}\log(p)} \big)j\log(p) \big]} + \frac{ e^{-\frac{n-j-|M_{o}|}{2} + j\log(p)} }{ \sqrt{\pi(n-j-|M_{o}|)} }.  \\
    & \le n^{\alpha} e^{-b_{n}\big[2n^{\alpha} - \frac{|M_{o}|}{2b_{n}} + \big(2 - \frac{1}{b_{n}} - \frac{1}{2b_{n}\log(p)} \big)(|M_{o}|+1)\log(p) \big]} \\
    & \ \ \ \ \ \ \ \ \ \ \ \ \ \ \ \ \ \ \ \ \ \ \ \ \ \ \ \ \ \ \ \ \ \ \ \ \ \ \ \ \ \ \ \ \ \ \ \ \ + \frac{ n^{\alpha}e^{-\frac{n-n^{\alpha}-|M_{o}|}{2} + n^{\alpha}\log(p)} }{ \sqrt{\pi(n-n^{\alpha}-|M_{o}|)} },  \\
    \end{split}
    \vspace{-.18cm}\]
    where $b_{n} := \frac{n-n^{\alpha}-|M_{o}|}{n-|M_{o}|-2} \to 1$.  Note that this bound vanishes by Condition \ref{condition2}.  Therefore, with probability exceeding one minus the above bound,
    \vspace{-.18cm}\[\small
    \begin{split}
    \Big(\frac{\RSS_{M_{o}}}{\RSS_{M}}\Big)^{\frac{n-j-1}{2}} & \le \Big(1 + \frac{4e^{2}(n^{\alpha} + j\log(p))}{\frac{n - j - 1}{2}}\Big)^{\frac{n-j-1}{2}} \\
    & \le e^{4e^{2}(n^{\alpha} + j\log(p))}, \\
    \end{split}
    \vspace{-.18cm}\]
    uniformly over all $M$ such that $|M_{o}| < |M| \le n^{\alpha}$.
\end{description}
$\hfill \blacksquare$

{\noindent \bf Proof of Theorem \ref{MainResult}.}
Without loss of generality suppose $\sigma_{M_{o}}^{0} = 1$, where $\sigma_{M_{o}}^{0}$ is the true but unknown error standard deviation.  For $j \in \{1,\dots,p\}$ define the following classes of subsets $\mathcal{M}_{j} := \{M \ne M_{o}: |M| = j\}$.  Recall that $\widehat{\sigma}_{M}^{2} := \RSS_{M} / (n - |M|)$.  It will first be shown that for any subset of covariates $M \ne M_{o}$ the ratio $\frac{r(M | Y)}{r(M_{o} | Y)}$ vanishes in probability for large $n$ and $p$.  Accordingly, for any $M \in \mathcal{M}_{j}$,
\vspace{-.18cm}\[\small
\begin{split}
\frac{r(M | Y)}{r(M_{o} | Y)} & = \pi^{\frac{j-|M_{o}|}{2}}\frac{\Gamma\Big(\frac{n-j}{2}\Big)}{\Gamma\Big(\frac{n-|M_{o}|}{2}\Big)} \frac{\RSS_{M_{o}}^{\frac{n-|M_{o}|-1}{2}}}{\RSS_{M}^{\frac{n-j-1}{2}}}\frac{\E(h(\beta_{M}))}{\E(h(\beta_{M_{o}}))} \\
& = \pi^{\frac{j-|M_{o}|}{2}}\frac{\Gamma\Big(\frac{n-j}{2}\Big)}{\Gamma\Big(\frac{n-|M_{o}|}{2}\Big)} \Big(\frac{\RSS_{M_{o}}}{\RSS_{M}}\Big)^{\frac{n-j-1}{2}}\RSS_{M_{o}}^{\frac{j-|M_{o}|}{2}}\frac{\E(h(\beta_{M}))}{\E(h(\beta_{M_{o}}))} \\
\end{split}
\vspace{-.18cm}\]

Before proceeding with the rest of the proof, a the following notation is needed.  $V_{1}$ and $V_{2}$ are as stated in Lemma \ref{RSSLemma}.  As in Theorem \ref{TrueModelCase}, define
\vspace{-.18cm}\[\small
V_{3} := \frac{3p_{M_{o}}\sigma_{M}\sqrt{\Lambda_{M_{o}}}}{\sqrt{\varepsilon\pi}}e^{-\frac{\varepsilon}{9\sigma^{2}_{M}\Lambda_{M_{o}}}},
\vspace{-.18cm}\]
and corresponding to Theorem \ref{BigModel}, define
\vspace{-.18cm}\[\small
V_{4} := \sum_{j=1}^{n^{\alpha}}\sum_{M\in \mathcal{M}_{j}}\frac{3|M|\sigma_{M}\sqrt{\Lambda_{M}}}{\sqrt{\varepsilon\pi}}e^{-\frac{\varepsilon}{9\sigma^{2}_{M}\Lambda_{M}}} \le \sum_{j=1}^{n^{\alpha}} \max_{\mathcal{M}_{j}} \frac{3|M|\sigma_{M}\sqrt{\Lambda_{M}}}{\sqrt{\varepsilon\pi}e^{\frac{\varepsilon}{9\sigma^{2}_{M}\Lambda_{M}} + j\log(p)} }
\vspace{-.18cm}\]
by bounding the binomial coefficient as in (\ref{binomCoefBehavior}).  Note that $V_{4}$ then vanishes by Condition \ref{condition4}.
Further, recall that $\RSS_{M_{o}} \sim \chi_{n-|M_{o}|}^{2}$, so by the Chernoff bound (evaluating the moment generating function at $1/4$),
\vspace{-.18cm}\[\small
P_{y}\big(\chi_{n-|M_{o}|}^{2} > 3(n - |M_{o}|)\big) \le \underbrace{ e^{-\frac{n-|M_{o}|}{4}} }_{\ds =: V_{5}}.
\vspace{-.18cm}\]
With this notation it is now possible to account for all of the uncertainty due to $Y$.

Accordingly, by Theorem \ref{TrueModelCase}, with probability exceeding $1 - V_{3} - V_{5}$,
\vspace{-.18cm}\begin{equation}\small\label{preliminaryBound}
\frac{r(M | Y)}{r(M_{o} | Y)} \le \frac{\Gamma\Big(\frac{n-j}{2}\Big)}{\Gamma\Big(\frac{n-|M_{o}|}{2}\Big)} \Big(\frac{\RSS_{M_{o}}}{\RSS_{M}}\Big)^{\frac{n-j-1}{2}}\frac{(3\pi(n-|M_{o}|))^{\frac{j-|M_{o}|}{2}}\E(h(\beta_{M}))}{1 - g_{1}(M_{o},n,p)},
\vspace{-.18cm}\end{equation}
where 
\vspace{-.18cm}\[\small
g_{1}(M,n,p) := \frac{2^{\frac{3}{2}}3j\widehat{\sigma}_{M}\sqrt{\Lambda_{M}}}{\sqrt{\pi\varepsilon}(1 - \frac{1}{n-j})}e^{-\frac{\varepsilon}{18\Lambda_{M}\widehat{\sigma}_{M}^{2}}}.
\vspace{-.18cm}\]
Further, fix $A_{1} \in (0,1)$, and by Condition \ref{condition4} choose $n$ and $p$ sufficiently large so that $\ds g_{1}(M_{o},n,p) < A_{1}$.

Suppose $j \le |M_{o}|$.  As in \cite{Jameson2013}, the ratio of gamma functions can be bounded by
\vspace{-.18cm}\[\small
\frac{\Gamma\Big(\frac{n-j}{2}\Big)}{\Gamma\Big(\frac{n-|M_{o}|}{2}\Big)} = \frac{\Gamma\Big(\frac{n-|M_{o}|}{2} + \frac{|M_{o}|-j}{2}\Big)}{\Gamma\Big(\frac{n-|M_{o}|}{2}\Big)} \le \Big(\frac{n-|M_{o}|}{2}\Big)\Big(\frac{n-j}{2}\Big)^{\frac{|M_{o}|-j}{2} - 1}.
\vspace{-.18cm}\]
Applying Lemma \ref{RSSLemma} to bound the ratio of residual sums of squares, and bounding the expectation by 1, with probability exceeding $1 - V_{1} - V_{3} - V_{5}$, (\ref{preliminaryBound}) implies
\vspace{-.18cm}\begin{equation}\small\label{smallBound}
\begin{split}
\frac{r(M | Y)}{r(M_{o} | Y)} & \le \Big(\frac{n-|M_{o}|}{2}\Big)\Big(\frac{n-j}{2}\Big)^{\frac{|M_{o}|-j}{2} - 1} \frac{ (3\pi(n-|M_{o}|))^{\frac{j-|M_{o}|}{2}}}{(1 - A_{1})e^{2\log(n)^{\gamma}\log(p)}} \\
& \le \frac{e^{-2\log(n)^{\gamma}\log(p)}}{1 - A_{1}}, \\
\end{split}
\vspace{-.18cm}\end{equation}
where the last inequality follows for all $n \ge \frac{|M_{o}|}{1 - \frac{1}{6\pi}}$.  Since $n >> |M_{o}|$, assume without loss of generality that $n$ is sufficiently large. 

Conversely, suppose $|M_{o}| < j \le n^{\alpha}$.  In this setting, as in \cite{Jameson2013}, the ratio of gamma functions can be bounded by
\vspace{-.18cm}\[\small
\left(\frac{\Gamma\Big(\frac{n-j}{2}\Big)}{\Gamma\Big(\frac{n-|M_{o}|}{2}\Big)}\right)^{-1} = \frac{\Gamma\Big(\frac{n-|M_{o}|}{2}\Big)}{\Gamma\Big(\frac{n-j}{2}\Big)} = \frac{\Gamma\Big(\frac{n-j}{2} + \frac{j - |M_{o}|}{2}\Big)}{\Gamma\Big(\frac{n-j}{2}\Big)} \ge \Big(\frac{n-j}{2} - 1\Big)^{\frac{j-|M_{o}|}{2}}.
\vspace{-.18cm}\]
Applying Theorem \ref{BigModel} to bound the expectation, and applying Lemma \ref{RSSLemma} to bound the ratio of residual sums of squares, with probability exceeding $1 - V_{2} - V_{3} - V_{4} - V_{5}$, (\ref{preliminaryBound}) implies
\vspace{-.18cm}\begin{equation}\small\label{bigBound}
\begin{split}
\frac{r(M | Y)}{r(M_{o} | Y)} & \le \frac{  e^{4e^{2}(n^{\alpha} + j\log(p))}   }{  \Big(\frac{n-j}{2} - 1\Big)^{\frac{j-|M_{o}|}{2}}   } (3\pi(n-|M_{o}|))^{\frac{j-|M_{o}|}{2}}\frac{g_{1}(M,n,p)}{1 - A_{1}} \\
& \le \frac{2^{\frac{3}{2}}3j\widehat{\sigma}_{M}\sqrt{\Lambda_{M}}  e^{-\frac{\varepsilon}{18\Lambda_{M}\widehat{\sigma}_{M}^{2}} + 4e^{2}(n^{\alpha} + j\log(p)) + \frac{j-|M_{o}|}{2}\log\Big(\frac{6\pi}{1-\frac{n^{\alpha}+2}{n}}\Big) }           }{  \sqrt{\pi\varepsilon}(1 - \frac{1}{n-j})(1 - A_{1})  } \\
\end{split}
\vspace{-.18cm}\end{equation}
Notice that Theorem \ref{BigModel} is being applied here with $n^{\alpha}$ in place of $c_{1}n$.  This can be done without loss of generality because $n^{\alpha}$ grows slower than $c_{1}n$, for any choices of $\alpha, c_{1} \in (0,1)$.

It can now be shown that $\sum_{j=1}^{n^{\alpha}}\sum_{M\in \mathcal{M}_{j}}\frac{r(M | Y)}{r(M_{o} | Y)}$ vanishes in probability for large $n$ and $p$.  Apply the bounds in (\ref{smallBound}) and (\ref{bigBound}) in the following argument.
\vspace{-.18cm}\[\small
\sum_{j=1}^{n^{\alpha}}\sum_{M\in \mathcal{M}_{j}}\frac{r(M | Y)}{r(M_{o} | Y)} \le \underbrace{\sum_{j=1}^{|M_{o}|}\binom{p}{j}\max_{M\in \mathcal{M}_{j}}\frac{r(M | Y)}{r(M_{o} | Y)}}_{\ds =: S_{1}} + \underbrace{\sum_{j=|M_{o}|+1}^{n^{\alpha}}\binom{p}{j}\max_{|M|=j}\frac{r(M | Y)}{r(M_{o} | Y)}}_{\ds =: S_{2}} 
\vspace{-.18cm}\]
Consider $S_{1}$ and $S_{2}$ separately.  

With probability exceeding $1 - V_{1} - V_{3} - V_{5}$,
\vspace{-.18cm}\[\small
S_{1} \le \frac{1}{1 - A_{1}}\sum_{j=1}^{|M_{o}|}e^{-2\log(n)^{\gamma}\log(p) + j\log(p)} \le \frac{|M_{o}|e^{-(2\log(n)^{\gamma} - |M_{o}|)\log(p)}}{1 - A_{1}}, 
\vspace{-.18cm}\]
by bounding the binomial coefficient as in (\ref{binomCoefBehavior}), and with probability exceeding $1 - V_{2} - V_{3} - V_{4} - V_{5}$,
\vspace{-.18cm}\[\small
\begin{split}
S_{2} & \le A_{2}\sum_{j=|M_{o}|+1}^{n^{\alpha}} \frac{  \frac{j}{1 - \frac{1}{n-j}} \max\limits_{|M|=j}\Big(\frac{\widehat{\sigma}_{M}^{2}\Lambda_{M}}{\varepsilon}\Big)^{\frac{1}{2}}  }{  e^{\min\limits_{|M|=j}\Big\{\frac{\varepsilon}{18\Lambda_{M}\widehat{\sigma}_{M}^{2}} - 4e^{2}(n^{\alpha} + j\log(p)) - \frac{j-|M_{o}|}{2}\log\Big(\frac{6\pi}{1-\frac{n^{\alpha}+2}{n}}\Big) - j\log(p) \Big\} }  } \\
& \le \frac{A_{2}}{n^{1 - \alpha}} \cdot \max\limits_{|M_{o}|<|M|\le n^{\alpha}}\Big(\frac{\widehat{\sigma}_{M}^{2}\Lambda_{M}}{\varepsilon}\Big)^{\frac{1}{2}}, \\
\end{split}
\vspace{-.18cm}\]
for some positive constant $A_{2}$.  The last inequality follows by Condition \ref{condition4}.

Thus, for sufficiently large $n$ and $p$, with probability exceeding $1 - V_{1} - V_{2} - V_{3} - V_{4} - V_{5}$,
\vspace{-.18cm}\[\small
\sum_{j=1}^{n^{\alpha}}\sum_{M\in \mathcal{M}_{j}}\frac{r(M | Y)}{r(M_{o} | Y)} \le \frac{|M_{o}|e^{-(2\log(n)^{\gamma} - |M_{o}|)\log(p)}}{1 - A_{1}}+  \frac{A_{2}}{n^{1 - \alpha}} \max\limits_{|M_{o}|<|M|\le n^{\alpha}}\Big(\frac{\widehat{\sigma}_{M}^{2}\Lambda_{M}}{\varepsilon}\Big)^{\frac{1}{2}}
\vspace{-.18cm}\]
which by Conditions \ref{condition1}-\ref{condition4} vanishes for large $n$ and $p$.  The proof is now complete by noticing that
\vspace{-.18cm}\[\small
r(M_{o}|Y) = \frac{r(M_{o}|Y)}{\sum_{j=1}^{n^{\alpha}}\sum_{M:|M|=j}r(M | Y)} = \frac{1}{1 + \sum_{j=1}^{n^{\alpha}}\sum_{\mathcal{M}_{j}}\frac{r(M | Y)}{r(M_{o}|Y)}}.
\vspace{-.18cm}\]
$\hfill \blacksquare$

\end{document}